\documentclass[fleqn,usenatbib]{mnras}
\usepackage{graphicx}	
\usepackage{amsmath}	
\usepackage{amssymb}	
\usepackage{apjfonts}
\usepackage{diagbox}
\usepackage{epsfig}
\usepackage{epstopdf}

\title[Single-pulse search approach]{A novel single-pulse search approach to detection of dispersed radio pulses using clustering and supervised machine learning}

\author[D. Pang et al.]{Di Pang$^{1,3}$\thanks{E-mail: dipang@mix.wvu.edu},
Katerina Goseva-Popstojanova$^{1,3}$\thanks{E-mail: Katerina.Goseva@mail.wvu.edu},
Thomas Devine$^{1,3}$
and Maura McLaughlin$^{2,3}$\thanks{E-mail: Maura.McLaughlin@mail.wvu.edu}
\\
$^1$Lane Department of Computer Science and Electrical Engineering, West Virginia University, PO Box 6201, Morgantown, WV 26506, USA\\
$^2$Department of Physics and Astronomy, West Virginia University, PO Box 6201, Morgantown, WV 26506, USA\\
$^3$Centre for Gravitational Waves and Cosmology, Chestnut Ridge Research Building, West Virginia University, Morgantown, WV 26505, USA
}

\date{Accepted 2018 July 17. Received 2018 July 13; in original form 2017 November 01}

\pubyear{2018}

\begin{document}
\label{firstpage}
\pagerange{\pageref{firstpage}--\pageref{lastpage}}
\maketitle

\begin{abstract}
We present a novel two-stage approach which combines unsupervised and supervised machine learning to automatically identify and classify single pulses in radio pulsar search data. 
In the first stage, we identify astrophysical pulse candidates in the data, which were derived from the Pulsar Arecibo L-Band Feed Array (PALFA) survey and contain 47,042 independent beams, as trial single-pulse event groups (SPEGs) by clustering single-pulse events and merging clusters that fall within the expected DM and time span of astrophysical pulses. 
We also present a new peak scoring algorithm, to identify astrophysical peaks in S/N versus DM curves. 
Furthermore, we group SPEGs detected at a consistent DM for they were likely emitted by the same source. 
In the second stage, we create a fully labelled benchmark data set by selecting a subset of data with SPEGs identified (using stage 1 procedures), their features extracted and individual SPEGs manually labelled, and then train classifiers using supervised machine learning. 
Next, using the best trained classifier, we automatically classify unlabelled SPEGs identified in the full data set.
To aid the examination of dim SPEGs, we develop an algorithm that searches for an underlying periodicity among grouped SPEGs.
The results showed that RandomForest with SMOTE treatment was the best learner, with a recall of 95.6\% and a false positive rate of 2.0\%.
In total, besides all 60 known pulsars from the benchmark data set, the model found 32 additional (i.e., not included in the benchmark data set) known pulsars, and several potential discoveries. 
\end{abstract}




\section{Introduction}
\label{sect:intro}
The study of radio pulsars provides us useful information on a wide range of topics in astronomy and physics, such as neutron star physics, general relativity, the Galactic gravitational potential and magnetic field, the interstellar medium, and so on \citep{2004hpa..book.....L}. Therefore, many pulsar surveys have been carried out aiming to discover more pulsars. 
It is estimated that there are over 100,000 detectable pulsars in the Milky Way \citep{Swiggum2014}. 
However, currently there are just over 2,600 known pulsars in the ATNF
catalog \citep{ATNF_catalog}. 

Pulsar search approaches are mainly divided into two categories: periodicity searches and single-pulse searches. While they both search 
the times series data for radio signals, 
periodicity searches look for periodic signals by first transforming the time series into the frequency domain using Fast Fourier Transforms (FFTs), and 
then folding the original time series at a period of interest to increase the signal-to-noise (S/N) ratio. 
In contrast, single-pulse searches look for bright, individual pulses without applying FFTs or folding \citep{Cordes2003}. 
Single-pulse searches are well suited for the discovery of isolated bursts that cannot be found in periodicity searches. Their application has led to the discoveries of Rotating Radio Transients (RRATs) \citep{McLaughlin2006} and Fast Radio Bursts (FRBs) \citep{Lorimer2007}, which consequently has renewed the interest in single-pulse searches \citep{Burke-Spolaor2010}.

Traditionally, in both pulsar search methods, pulsars are discovered through manual inspection of the candidates' diagnostic plots  \citep{Deneva2009, Burke-Spolaor2010}. 
However, due to the large number of observations and therefore diagnostic plots, the manual inspection is time-consuming and tedious. 
For example, the data set analysed in this paper was derived from the Pulsar Arecibo L-band Feed Array (PALFA) survey and it contained 47,042 independent beams.
Furthermore, in single-pulse searches, it is often advantageous to produce multiple diagnostic plots for different dispersion measure (DM\footnote{DM is defined as the integrated free-electron column density along the line of sight in units of $\textrm{pc}\ {\textrm{cm}^{-3}} $.}) ranges. 
For four such ranges (i.e., 0-30, 20-110, 100-310, 300-1000+ $\textrm{pc}\ {\textrm{cm}^{-3}} $) used in the PALFA survey,
161,888 diagnostic plots must be generated and manually inspected. 
Today, manual inspection stills plays a vital role in pulsar discovery, but this must change in the near future as wide-field many-pixel surveys come online.

A more efficient pulsar search approach is to first rank the candidates and then based on their astrophysical features only examine the best candidates manually.
A representative approach of this type is RRATtrap, proposed by Karako et al. (\citeyear{Karako-Argaman2015}).
In their algorithm, single-pulse events at similar DMs and times were first placed into groups. Each group was then assigned a rank using a set of predetermined rules related to its astrophysical features. 
Finally, only diagnostic plots that included groups with ranks higher than a threshold were inspected. 
This approach was able to reduce the number of diagnostic plots that required manual examination by an order of magnitude. 
Similar sorting algorithm have been applied in periodicity searches as well \citep{Keith2009}.
However, because Radio Frequency Interference (RFI) and true pulsar signals can be similar in appearance and the shapes of pulsar signals can vary significantly, it is difficult to find a set of efficient, predetermined rules that universally work.

Machine learning is well-suited for solving complex problems like pulsar searches. Not only does machine learning enable learning without being explicitly programmed, the performance of intelligent algorithms also improves when exposed to more data. Ideally, with machine learning, the candidates can be truly automatically classified. 
While machine learning approaches have been extensively applied 
in periodicity searches \citep{Bethapudi2018}, little research has explored their application to single-pulse searches. 
Devine et al. presented the first machine learning approach to single-pulse searches \citep{Devine2016}. 
This approach first identified pulsar signal candidates by finding dispersed pulse groups (DPGs) via inspecting how their S/N varied with DM using a peak-identification algorithm called Recursive Algorithm for Peak IDentification (RAPID). It then classified DPGs automatically using supervised machine learning. This approach greatly reduced the number of candidates that needed to be examined manually. 
However, the implementation only made use of the composite S/N versus DM subplot for an entire observation, meaning that dimmer pulses could be hidden by brighter pulses or RFI and hence overlooked.

In order to successfully apply machine learning in single-pulse searches, several challenges have to be addressed. 
The first challenge is related to ubiquitous RFI signals, which vary drastically in properties and brightness. 
Furthermore, some RFI signals look like astrophysical pulses, making it difficult to distinguish them. 
Second, the automatic approach should be able to detect a variety of pulses. This is because astrophysical pulses can vary significantly in brightness, width, and shape, sometimes even for pulses from the same source. 
Third, data processing techniques lead to additional challenges.
For example, variable spacing of trial DMs \citep{Cordes2003} and clipping \citep{RansomPhDT} can change the appearance of pulses in the S/N versus DM space, as well as in the DM versus time space, making their identification more complicated. 
Lastly, the large number of candidates calls for an efficient machine learning approach with low false positive rate and even lower false negative rate. 
In summary, there is still much to improve in automatic single-pulse search approaches.

In this paper, we present a novel single-pulse search approach which combines clustering and supervised machine learning methods to automatically identify and classify radio pulses.
Our approach addresses, 
for the first time,
several challenges caused by either the astrophysical properties of radio signals or the insufficiencies of current data processing techniques. 
This approach consists of two stages. 
In the first stage, our Single-Pulse Event Group IDentification (SPEGID) algorithm first identifies
astrophysical pulse candidates as trial
single-pulse event groups (SPEGs) by clustering (i.e., unsupervised learning) of related 
trial single-pulse events\footnote{Trial single-pulse events are obtained when we search for pulses across a range of trial DM values in the DM versus time space, and they are called single-pulse events for short in the rest of this paper.}
and merging of clusters, then calculates the peak scores of SPEGs in the S/N versus DM space, and finally groups SPEGs that appear at a consistent DM.
The output of the first stage is the identified SPEGs (i.e., astrophysical pulse candidates), which then need to be classified as pulsar and non-pulsar events. Note that by removing noise events and some RFI, the first stage significantly denoises the DM versus time subplot of the diagnostic plots. 
Such denoising effect makes the astrophysical pulses more noticeable and, when used in combination with the SPEGs' peak scores, can significantly reduce the effort needed for manual inspection of the diagnostic plots. 
However, when the number of diagnostic plots is large, inspecting them
manually can still be a slow and tedious process. Therefore, instead of manual classification, in the second stage (which was built on our prior work \citep{Devine2016}), we use supervised machine learning techniques to automatically classify SPEGs as pulsars and non-pulsars.  

The data used in this paper were derived from the PALFA survey, which is an ongoing long-term pulsar survey of the Galactic plane that started in 2004 \citep{Cordes2006}. 
This survey operates at a central frequency of 1.4 GHz using the ALFA receiver.
The observation lengths are 268 s for the inner Galaxy pointings and 180 s for outer Galaxy pointings \citep{Lazarus2015}.
ALFA is a seven-beam feed array arranged in an hexagonal pattern, with the centre pixel (beam 0) being surrounded by a ring of 6 pixels (beams 1-6).\footnote{In the rest of the paper, we use beam instead of pixel to follow the convention of related works \citep{Spitler2014, Swiggum2014}.}
This means each ALFA pointing includes seven distinct beam positions and thus during the observation seven independent beams are collected simultaneously \citep{Spitler2014, Swiggum2014}.
The beams are separated from each other by approximately one beam-width on the sky.
Therefore, with each beam covering $3\arcmin.5$ (FWHM), the receiver has a combined power pattern of approximately $24\arcmin\times26\arcmin$, which makes it well suited for survey observations
\citep{Deneva2009}. 
Moreover, the 1.4 $\textrm{GHz}$
operating frequency is suitable for pulsar searching of the Galactic plane.
In this paper, we analysed 47,042 independent beams which were observed between March 2009 to May 2015. This includes 33,536 unique positions as some positions were observed multiple times. Additionally, not all pointings had all seven beams included due to beams missing in the available data.

Note that, while this paper presents only the results based on data from the PALFA survey, by changing a small number of (or perhaps no) parameters, SPEGID can be applied to data from other surveys that may have different observational setups. For details, see Section~\ref{sect:approach}.

The rest of the paper is organized as follows. Section~\ref{sect:background} provides more background on single-pulse searches. Our approach to SPEG identification and classification is described in Section~\ref{sect:approach} and the results of our experiments are presented in Section~\ref{sect:results}. The discussion and implications are given in Section~\ref{sect:discussion}. We present the conclusions in Section~\ref{sect:conclusion}.

\section{Background on Single-pulse Searches}
\label{sect:background}

Pulsar radio signals travel through interstellar medium (ISM) before they are detected by radio receivers on Earth. 
After data collection, the original time series data are processed using a procedure that includes dedispersion, matched filtering, and thresholding \citep{Cordes2003}.
Dispersion refers to the phenomenon that, because of the interaction between signals and ISM, lower frequency components of signals arrive later than the higher frequency components.
Hence the effect of dispersion has to be removed by dedispersion before the diagnostic plots are made \citep{2004hpa..book.....L}.
Besides frequency, the magnitude of the delay also depends on DM.
However, at the time of dedispersion, the true DM value of the source is unknown, therefore we must search for pulses across a range of trial DMs.

Typically, the time series for a particular DM is downsampled multiple times and re-searched, with the pulse reaching its highest S/N when the effective sample time is closest to the width of the pulse.
In different DM channels, the S/N decreases as the trial DM deviates from the true DM \citep{Cordes2003}.
As a result, a single pulse is typically detected over several to many adjacent DM channels.
We call these neighboring detections single-pulse events. 
Each single-pulse event has five recorded properties: S/N, time, DM, time index, and downsampling factor.  
RFI signals are also detected as a group of single-pulse events that cluster in the DM versus time 
space and mostly (but not always) have S/N peaking
at zero DM. 
In contrast, random noise usually shows up as an individual event.

\begin{figure}
	\centering
	\includegraphics[width=\columnwidth]{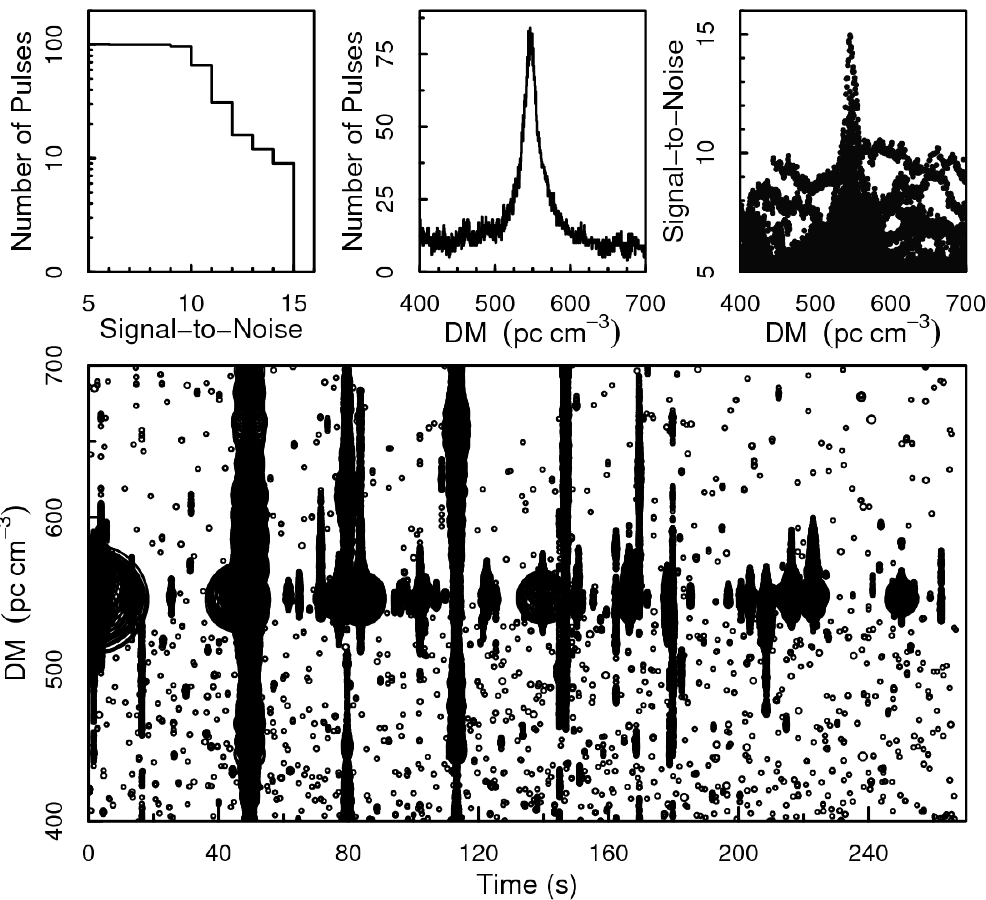}
	\caption{Known pulsar J1901+0355 detected on MJD 56245 in the PALFA survey. The four subplots include (clockwise from top left): 
		a histogram of the number of single-pulse events versus S/N, a histogram of the number of single-pulse events versus DM, a scatter plot of S/N versus DM and a scatter plot of the DM versus time for each single-pulse event whose is proportional to its S/N.}
	\label{fig:typical_pulsar}
\end{figure}

An example of a commonly used diagnostic plot is shown in Fig.~\ref{fig:typical_pulsar}, which comprises four subplots: a histogram of the number of single-pulse events versus S/N (top left), a histogram of the number of single-pulse events versus DM (top middle), a scatter plot of S/N versus DM (top right) and a scatter plot of DM versus time for each single-pulse event (bottom). 
Typically a combination of information shown in these subplots is used to determine which signals are astrophysical.

\section{Our Proposed Machine Learning Approach}
\label{sect:approach}
\begin{figure}
	\centering\includegraphics[width=\columnwidth]{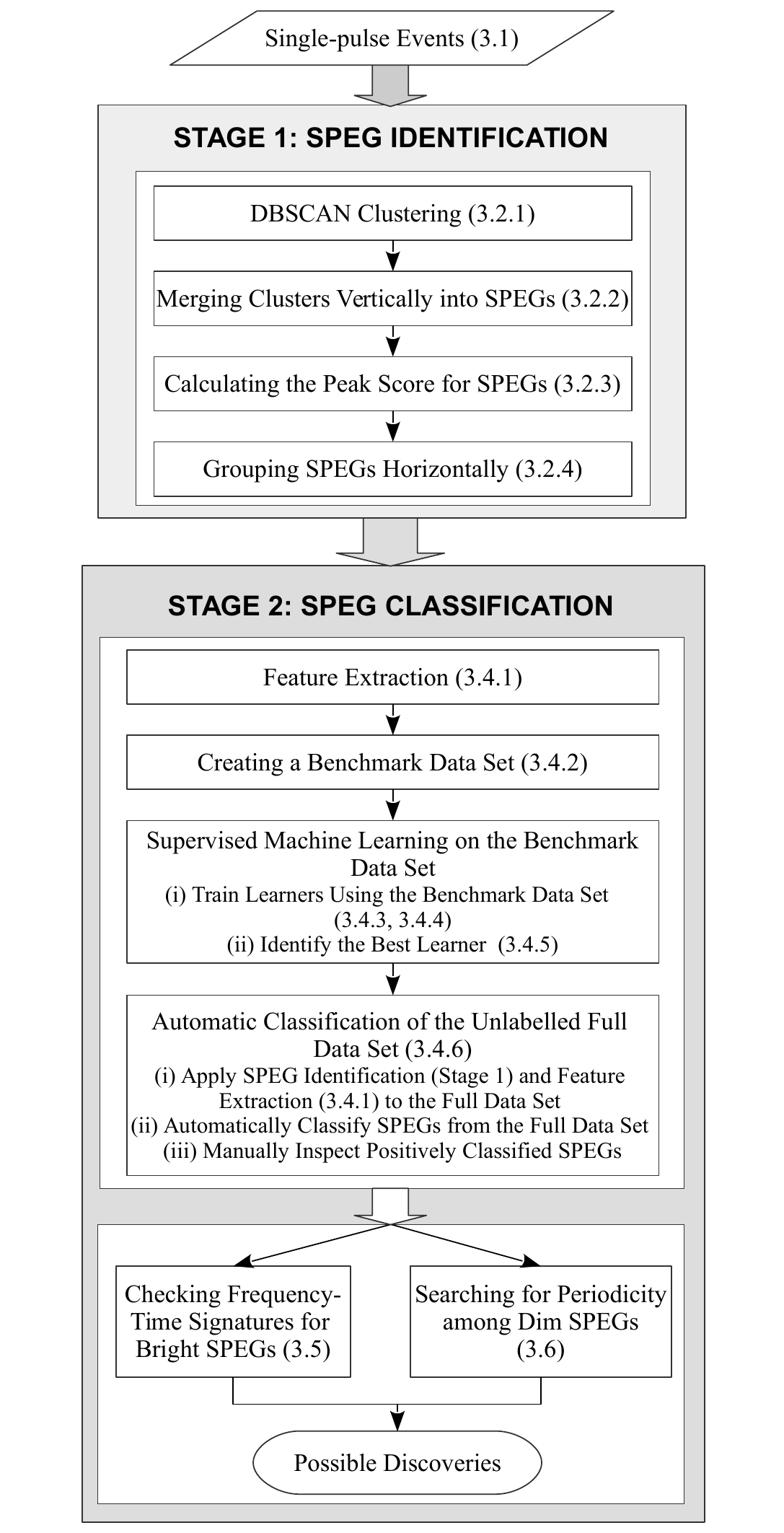}
	\caption{The outline of our proposed approach, with numbers in parentheses indicating the corresponding section in which the details are described.}
	\label{fig:flowchart}
\end{figure}

Fig.~\ref{fig:flowchart} shows the outline of our proposed two-stage approach in a flowchart, with numbers in parentheses indicating the corresponding section in which the details are described.
Briefly speaking, the first stage of our approach, namely SPEGID, includes the identification of pulse candidates as SPEGs that are made up of related single-pulse events, the characterisation of the S/N versus DM curve of SPEGs and the recognition of the association among them.
One benefit of stage 1 is that it removes most noise events and some RFI and, therefore, significantly denoises the DM versus time space of the diagnostic plots. Consequently, the manual inspection of the denoised diagnostic subplots (with SPEGs identified) would require much less effort compared to the original DM versus time subplots. 
However, when the number of diagnostic plots is large, manual inspection would still be tedious and time-consuming.
Therefore, instead of manual classification, we use the output from the first stage as input to the second stage to automatically classify the identified SPEGs as pulsar and non-pulsar events. 

The novel aspects of SPEGID include the following: 
\begin{enumerate}
	\item We successfully identified pulse candidates as SPEGs in the DM versus time subplots by first applying Density Based Spatial Clustering of Applications with Noise (DBSCAN) (i.e., unsupervised learning) \citep{Ester1996} on single-pulse events and then merging the clusters based on the relations among their S/N, width, DM offset (i.e., the absolute difference between the trial DM and the true DM) \citep{Cordes2003} and time drift.
	Note that clustering and merging of clusters are two integral parts for the identification of pulse candidates as SPEGs.
	\item We developed a new peak scoring algorithm which is capable of identifying the peak in the S/N versus DM curve of pulses/SPEGs with different brightness, width, and shapes. This algorithm also accounts for variable spacing of trial DMs.
	\item We also grouped SPEGs that appeared 
	at a consistent DM but different times together as these groups would have a higher probability of having an astrophysical origin.
	\item We extended the feature set to characterise both individual SPEGs and the SPEG groups, and as in our prior work \citep{Devine2016} used supervised machine learning algorithms in combination with treatment for imbalanced data to automatically classify SPEGs as pulsars and non-pulsars.   
	\item Last but not least, we calculated the probabilities of a group of (three to five) purely randomly distributed single-pulse events that were found to have an underlying periodicity. 
	Additionally, we confirmed the astrophysical origin of dim SPEG candidates by searching for an underlying periodicity among them, which helped to differentiate dim SPEGs/pulses from RFI and noise SPEGs. 
\end{enumerate}

The code of SPEGID was implemented in \textsc{python 2.7} and made available in the Astrophysics Source Code Library.\footnote{\href{http://ascl.net/1807.014}{http://ascl.net/1807.014}}

\subsection{Single-pulse events}
\label{sect:spe}
The single-pulse search data set used in this paper was processed by the \textsc{presto} code $single\_pulse\_search.py$
\citep{RansomPhDT},
in which the spacing of the trial DM values (that were used to dedisperse the data) increases at higher DMs.
After matched filtering, only single-pulse events with S/N above 5 were recorded.
\begin{figure}
	\centering\includegraphics[width=\columnwidth]{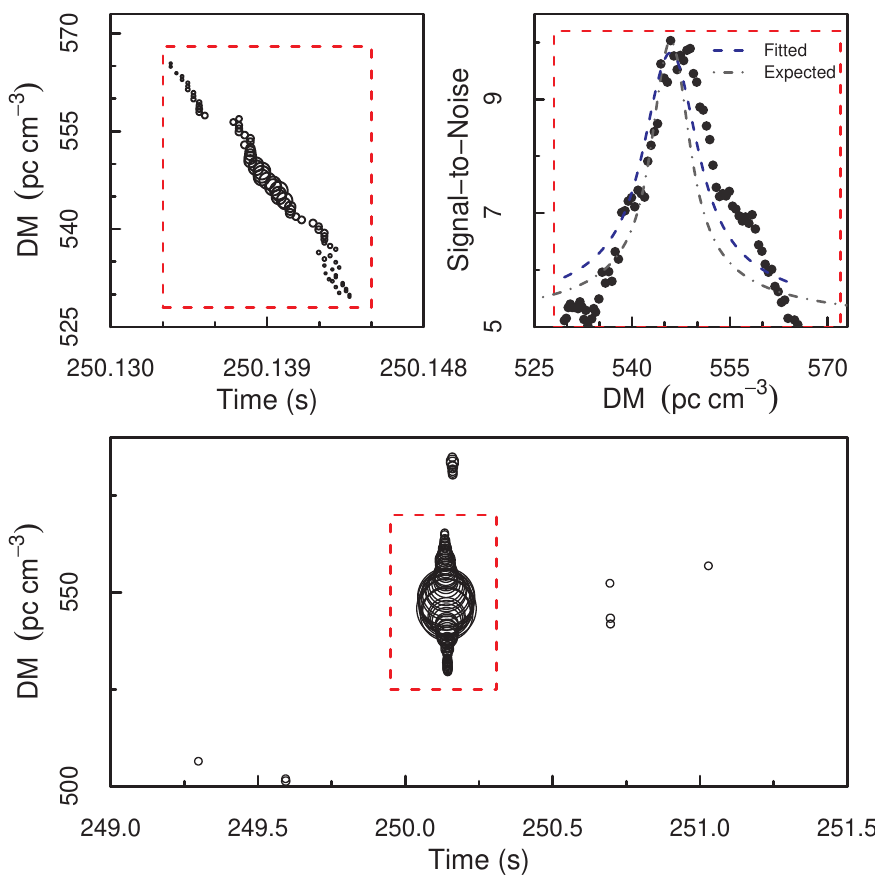}
	\caption{A typical astrophysical pulse detected as of a group of single-pulse events. The bottom and the top left subplots show its appearance in the DM versus time space, and the top right subplot show its S/N versus DM curve. Note that in both the top left and the bottom subplots,
		the single-pulse events are both plotted proportionally to S/N but with different scaling to show the single-pulse events more clearly in the top left subplot. In the top right subplot, we also show both the fitted and expected S/N decline that are caused by DM offset. Note that they are both fairly close to the observed S/N decline, and to each other as well.}
	\label{fig:typical_pulse}
\end{figure}
Fig.~\ref{fig:typical_pulse} shows a typical astrophysical pulse detected as a group of related single-pulse events. 
As it can be seen in the top left zoomed DM versus time subplot, there is a slight variance in time between neighboring single-pulse events.
For this reason, we must also understand the expected spread of single-pulse events in time in order to group them correctly. 
In fact, for two adjacent single-pulse events separated by one trial DM step, 
this difference in time is typically one time sample, 
given the way that DM channels are spaced and downsampling factors are chosen (see Section~\ref{sect:DBSCAN}). 
Meanwhile, as shown in the top right S/N versus DM subplot, the observed  $\textrm{S/N}$ of the pulse decreases as the DM offset ($\delta \textrm{DM}$) from the true DM increases. This relation is characterised by equation~(\ref{eq:SNR_DM}) \citep{Cordes2003}:
\begin{equation}
	\frac{\textrm{S}(\delta \textrm{DM})}{\textrm{S}} =  \frac{\sqrt{\pi}}{2}{\zeta^{-1}}{\textrm{erf} \zeta},
	\label{eq:SNR_DM}
\end{equation}
where ${\textrm{S}(\delta \textrm{DM})} / {\textrm{S}}$ is the ratio of observed $\textrm{S/N}$ to true peak $\textrm{S/N}$, erf is the error function, and $\zeta$ is given by
\begin{equation}
	{\zeta} = {6.91 \times 10^{-3}\delta{\textrm{DM}}}{\frac{\Delta \nu}{W \nu^{3}}}.
	\label{eq:SNR_DM_erf}
\end{equation}
In equation~(\ref{eq:SNR_DM_erf}), $\Delta \nu$ is the total bandwidth in MHz, $W$ is the pulse width (FWHM) in milliseconds, and $\nu$ is the central observing frequency in GHz.
Note that equations~(\ref{eq:SNR_DM}) and~(\ref{eq:SNR_DM_erf}) assume a Gaussian pulse. With these two equations, given the S/N and the width $W$ of an astrophysical pulse, we can calculate the expected threshold for the DM offset ($\delta \textrm{DM}$) that will make the observed S/N decrease to the noise level (i.e., S/N = 5).
An example is shown in the top right subplot in Fig.~\ref{fig:typical_pulse}.
In this subplot, we show both the expected and fitted S/N decline.
In both cases, the DM at which the S/N peaks is considered as the true DM. 
We calculated the expected S/N decline using the peak S/N and the pulse width obtained from matched filtering, which is a close approximation to the actual pulse width.
On the other hand, to calculate the fitted S/N decline, we first used a non-linear least squares (NLS) regression to obtain the fitted peak S/N and $W$ of the pulse and then calculated the fitted S/N decline using the same equations.
It can be seen that the expected S/N decline is fairly close to the fitted values even though the observed peak S/N and the approximate pulse width were used. 
Furthermore, by assuming a time drift of one sample for each DM step,
we can also find the expected time extent of this pulse.
In fact, we use this idea to merge clusters that likely originate from the same pulse, as described in Section ~\ref{sect:merging}.

\subsection{Identification of SPEGs}
\label{sect:identification_stage}

\subsubsection{DBSCAN clustering}
\label{sect:DBSCAN}
Clustering is an unsupervised learning method that involves grouping data objects with similar properties together \citep{Han:2011:DMC:1972541}. 
It is very useful in finding previously unknown groups in a data set. 
The similarity (and therefore, dissimilarity) between data objects is usually determined by the distance measure derived from their attributes. 
Clustering algorithms are divided into two categories: partitioning (non-hierarchical) and hierarchical.
While partitioning methods separate data objects into a number of clusters, hierarchical methods link the cluster pairs successively and thus form a nested hierarchy among clusters \citep{Rasmussen1992}.  
\begin{figure}
	\centering\includegraphics[width=\columnwidth]{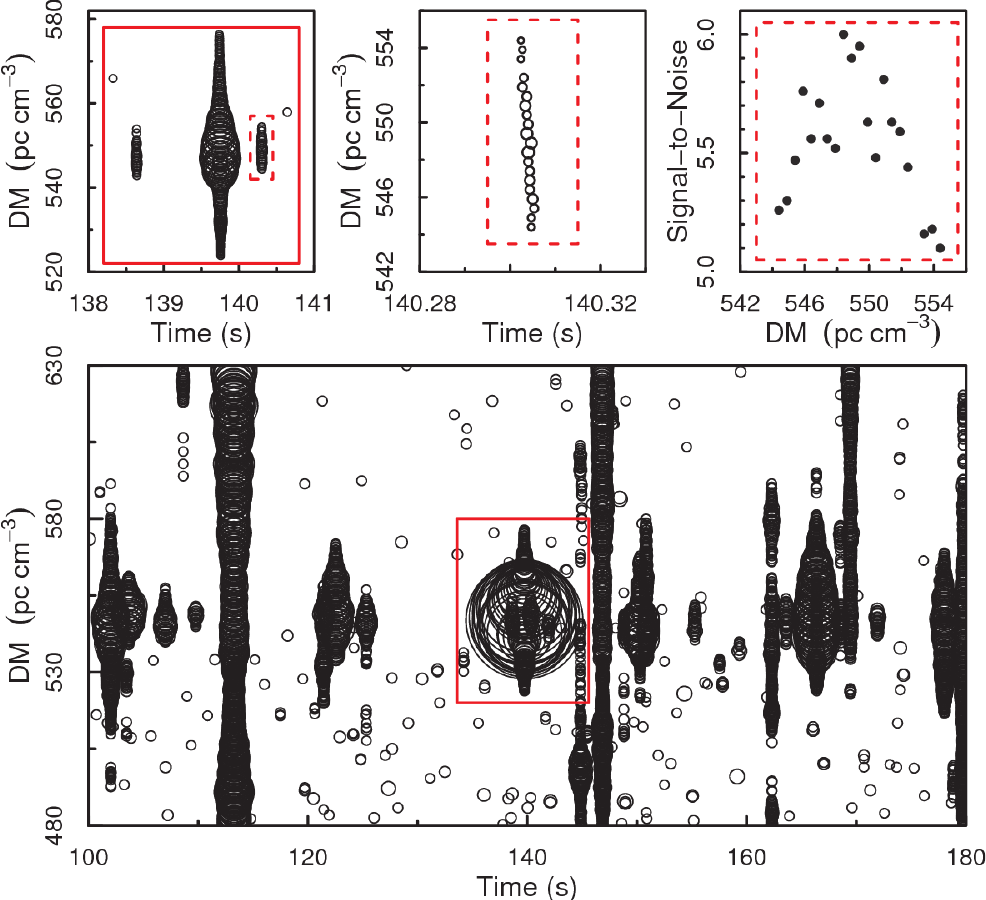}
	\caption{ A zoomed-in part of the diagnostic plot shown in Fig.~\ref{fig:typical_pulsar}. The region surrounded by the solid rectangle in the bottom subplot is further enlarged to show the details in the top left subplot, in which three clusters of single-pulse events can be seen. The smaller cluster on the right (surrounded by a dashed rectangle) is zoomed in and shown in the top middle and the top right subplots. This cluster is used to demonstrate DBSCAN clustering in Fig.~ \ref{fig:DBSCAN}.}
	\label{fig:zoomed_pulses}
\end{figure} 
Since related single-pulse events are typically close in DM and time,  partitioning clustering methods can be used to identify pulse candidates and separate them from each other.
Specifically, we applied DBSCAN clustering \citep{Ester1996} in the DM versus time space and thus partitioned the single-pulse events into signals and noise by measuring their density. We demonstrate below how DBSCAN clustering works on single-pulse events using one astrophysical pulse from Fig.~\ref{fig:zoomed_pulses} (which displays part of the diagnostic plot shown in Fig.~\ref{fig:typical_pulsar}) as an example. For this work, we used $scikit-learn$'s implementation of DBSCAN clustering in \textsc{python} 2.7 \citep{Scikit-learn}.

DBSCAN clustering requires two parameters: the density threshold ($MinPts$) of a core object and the radius ($\epsilon$) of its neighborhood. If an object has at least $MinPts$ of adjacent objects in its $\epsilon$-neighborhood, then this object is considered a core object. 
Core objects and their neighborhood are considered as dense regions and form clusters.
\begin{figure}
	\centering\includegraphics[width=\columnwidth]{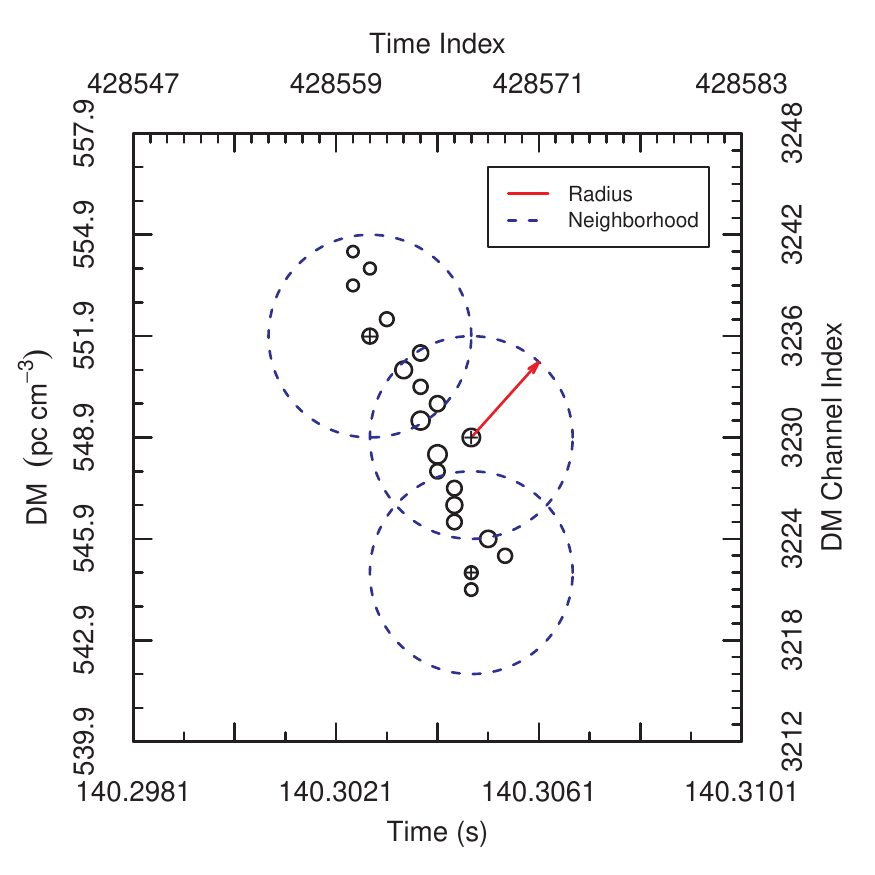}
	\caption{An example of DBSCAN clustering applied on the cluster from Fig.~\ref{fig:zoomed_pulses}. Note that instead of the actual time and DM, time and DM channel indices were used to calculate the distance between single-pulse events. 
		In this example, all single-pulse events are core objects in DBSCAN clustering and the radius of their neighborhood is small enough to avoid including any noise event.
		Because their neighborhood areas overlap with one another, all single-pulse events are joined together to form one single cluster.
		It can also be seen for two adjacent single-pulse events, a unit increase in DM channel index roughly corresponds to a unit decrease of time index.}
	\label{fig:DBSCAN}
\end{figure}
If the dense regions of two core objects overlap, they are joined together to form one single cluster. 
On the other hand, objects that are neither core objects themselves, nor fall into the neighborhood of any core object, are classified as noise \citep{Ester1996}.  
It should be emphasized that although the shape of the neighborhood is controlled by the distance metric between objects (for example, if Euclidean distance is used, the shape of the neighborhood will be circular), the shape of the clusters (as amalgamated core points with their neighborhood) can be arbitrary.


We chose to use Euclidean distance in our application, with time and DM channel indices as our dimensionless units (see Fig.~\ref{fig:DBSCAN}). (Even though other distance metrics could be used, they do not seem to have any particular advantage). Fig.~\ref{fig:DBSCAN} shows the small cluster of single-pulse events from the top left subplot of Fig.~\ref{fig:zoomed_pulses}. A closer examination shows that for two adjacent single-pulse events, as the DM increases by one DM spacing, the time decreases by approximately one sample. This is an artefact of the dedispersion in which higher DM channels have later start times; these pulses actually arrive at the same time in different DM channels. Our distance metric will allow for these slopes in the DM versus time plane.

The effect of DBSCAN clustering on single-pulse events is shown in Fig.~\ref{fig:DBSCAN}, using a density threshold $MinPts = 5$ and a radius $\epsilon = 6$ for the neighborhood. It can be seen that the single-pulses events successfully form a single cluster without including any noise event. 
Next, we describe the reasons behind choosing these particular values for the parameters $MinPts$ and $\epsilon$. 

\textit{$MinPts$ value selection}.
Unlike the bright, wide pulses which are detectable in many DM channels, narrow and/or dim pulses can only be detected in a few DM channels (especially when the DM spacing is large).
In extreme cases, we have observed pulses having less than 2 $\textrm{pc}\ {\textrm{cm}^{-3}} $ span in the DM domain. Such pulses would be missed by DBSCAN clustering if the DM spacing is large.
One way to solve this is to use denser DM spacing in the data processing step of the pipeline.
In the absence of that, our current best strategy is to choose a small $MinPts$ value, so that we can detect most pulses when DM spacing is small and as many pulses as possible when DM spacing becomes larger.
For this reason, we used $5$ as the optimal value for $MinPts$ so we could identity as many narrow and/or dim pulses as possible. Note that it would be difficult to confirm the expected S/N versus DM shape with less than five events. 

\textit{$\epsilon$ value selection.} Ideally speaking, if the single-pulse events form a diagonal line in the DM versus time space, we should be able to find 5 single-pulse events within a $2\sqrt{2}$ radius of a core object. 
However, because the same DM and time spacing was used for a range of DMs, the 1-to-1 relation between time and DM channel indices is only approximate for the entire DM range.
As shown in Fig.~\ref{fig:DBSCAN}, single-pulse events oftentimes do not form a diagonal line. 
Therefore, it is often observed that the maximum time drift between two adjacent single-pulse events is much larger than one time sample, especially in bright and wide pulses that are made of many single-pulse events. 
Due to possible large time difference and/or the occasional non-detection of single-pulse events in certain DM channels, we found that $\epsilon =6$ was optimal for discriminating pulses from other pulses and/or noise, while still allowing for larger than expected distances.

Generally speaking, the parameters of DBSCAN clustering $MinPts$ and $\epsilon$ can be modified to suit surveys with different observational setups and/or different research goals. For example, if the predetermined DM spacing is really dense (or in a less likely case, if narrow pulses are not of interest), the density threshold can be increased in order to filter out noise clusters that are made up of less than $MinPts$ single-pulse events. If a larger $MinPts$ is selected, a larger radius $\epsilon$ should be used accordingly.

For beams in which consistent RFI is detected during the observation duration or many pulses from a very bright pulsar are detected, many single-pulse events will be generated.
When DBSCAN clustering is applied to such beams, a large number of single-pulse event clusters will usually be formed.
Investigating all these clusters not only requires significant computing power, it is also unnecessary.
This is because in the former case, single-pulse event clusters with low maximum S/N are more likely from RFI than real astrophysical signals, whereas in the latter case, the identification of those bright pulses is sufficient for the detection of the pulsar. 
In other words, when the cluster density is high, we only need to consider those relatively bright clusters.
As a result, 
after running DBSCAN clustering, we calculated the cluster density as the number of clusters divided by the product of the number of trial DM channels ($\textrm{N}_\textrm{DM}$) and the observation duration ($t_\textrm{obs}$):
\begin{equation}
	\textrm{ClusterDensity} =  \frac{\textrm{N}_\textrm{cluster}}{\textrm{N}_\textrm{DM} \times t_\textrm{obs}} .
	\label{eq:cluster_density}
\end{equation}
Note that the cluster density of a beam is normalised over the observation duration hence it is independent of the observation duration. Such statement could not be made for the number of clusters as it is not uncommon for two observations to have different durations (even within the same survey).
Specifically, we define the clusters to be bright or dim as follows:
\begin{enumerate}
	\item If the cluster density is less than 0.01 per channel per second, all clusters (with maximum S/N $\ge$ 5) are further investigated. Clusters with maximum S/N $\ge$ 6 are considered as bright clusters, and the rest are considered as dim clusters;
	\item If the cluster density is between 0.01 and 0.02 per channel per second, only clusters with maximum S/N $\ge$ 5.5 are further investigated. Clusters with maximum S/N $\ge$ 6.5 are considered as bright clusters, and the rest are considered as dim clusters;
	\item If the cluster density is greater than 0.02 per channel per second, only clusters with maximum S/N $\ge$ 6 are further investigated. Clusters with maximum S/N $\ge$ 7 are considered as bright clusters, and the rest are considered as dim clusters.
\end{enumerate}
Considering that a typical PALFA observation has an observation length of 268 s, and the number of trial DM channels is often greater than 5,000, we only ignore single-pulse event clusters with peak S/N $<5.5$ when the total number of clusters exceeds 13,400 ($268 \times 5,000 \times 0.01$), and clusters with peak S/N $<6$ when the total number of clusters exceeds 26,800 ($268 \times 5,000 \times 0.02$). 
Our preliminary experimentation showed that: 
(i) For the majority of beams, usually much fewer clusters were found, therefore all clusters would be further investigated; 
(ii) For beams in which consistent RFI or very bright pulsars were observed, ignoring the dim clusters helped to decrease the running time of our algorithms significantly (which was necessary due to the large number of beams in our data set), yet minimised the possibility of missing potential pulsar discoveries. 
Note that because cluster density is independent of the observation duration and we selected fairly conservative threshold values for the sake of reasonable computation time, equation~(\ref{eq:cluster_density}) and the threshold values given above can be used for other surveys with different observation lengths.

The definition of bright clusters will be referred to later, when clusters are merged into SPEGs in Section~\ref{sect:merging}. SPEGs that contain at least one bright cluster will be considered as bright SPEGs and used to form SPEG groups in Section~\ref{sect:grouping}. 

\subsubsection{Merging clusters vertically into SPEGs}
\label{sect:merging}
By applying DBSCAN clustering, we are able to separate pulses from other pulses and/or noise.
However, instead of grouping all single-pulse events from the same astrophysical pulse into one SPEG, DBSCAN clustering (with the parameters given in Section~\ref{sect:DBSCAN}) sometimes resulted in ``broken'' SPEGs (i.e., single-pulse events from one astrophysical pulse were grouped into multiple clusters). This was because of the unusually large distance between adjacent single-pulse events, and is mainly due to the following two reasons.

\begin{figure}
	\centering\includegraphics[width=\columnwidth]{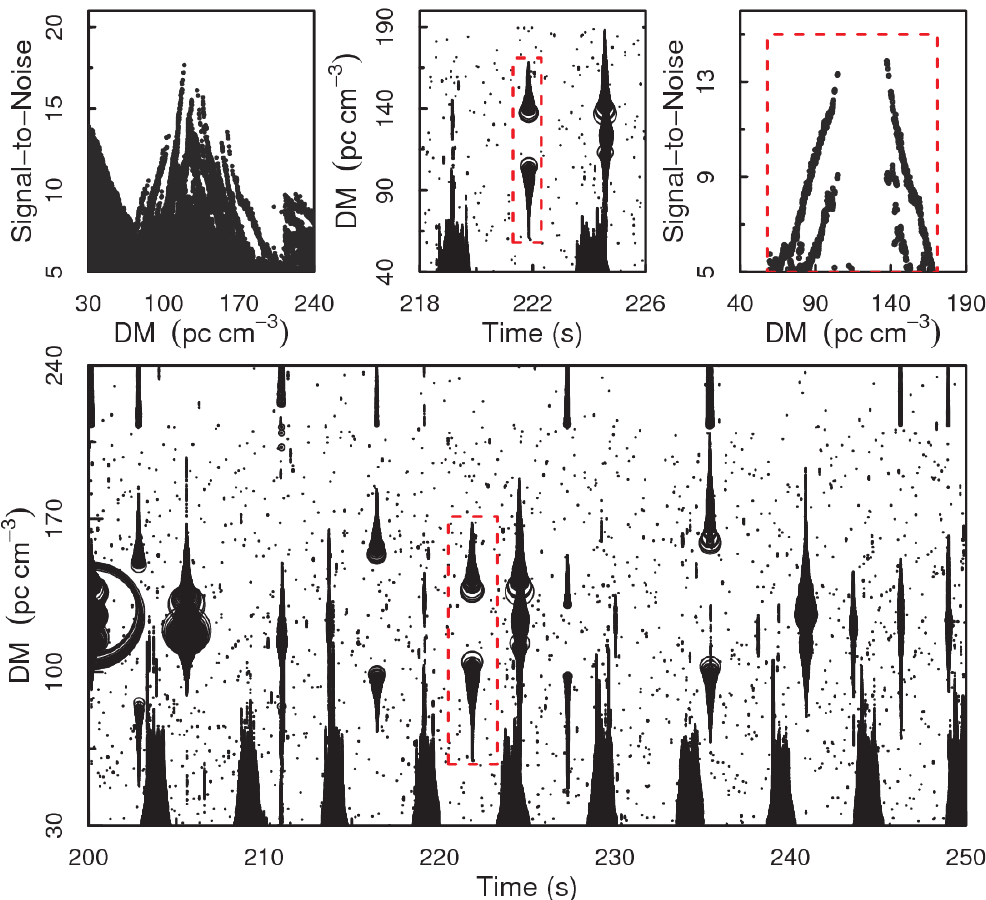}
	\caption{Known pulsar J1910+0714 detected on MJD 56663 in the PALFA survey. Note that there are many clipped pulses. The top left subplot and the bottom subplot show the S/N versus DM space, and the DM versus time space of this whole section respectively; and the other two subplots show those regions around the clipped pulse of interest that is surrounded by the red dashed line rectangle.}
	\label{fig:clipped_pulses}
\end{figure}

The first reason that can cause a larger than expected distance is clipping. 
Often, in pulsar search algorithms, samples with very high intensity are ``clipped'', or replaced by the median values of the time series. This can affect both bright pulses and RFI, and is used by the \textsc{presto} $prepsubband$ command used in the PALFA pipeline.
As a result, we often observed two different types of pulses in single-pulse search output: regular (non-clipped) pulses (see Fig.~\ref{fig:typical_pulse}) and clipped pulses (shown in Fig.~\ref{fig:clipped_pulses}).
As can be seen in Fig.~\ref{fig:clipped_pulses}, clipping has removed the brightest single-pulse events from the DM versus time subplot and, correspondingly, peaks from the S/N versus DM subplot, which leads to ``broken'' astrophysical pulses. 
If the broken parts of a pulse are not merged together, no peak-like shape would be detected in the S/N versus DM subplot and the signal would be interpreted as RFI. 
This is especially problematic for FRBs or RRATs that emit a small number of bright pulses during an observation.

\begin{figure}
	\centering\includegraphics[width=\columnwidth]{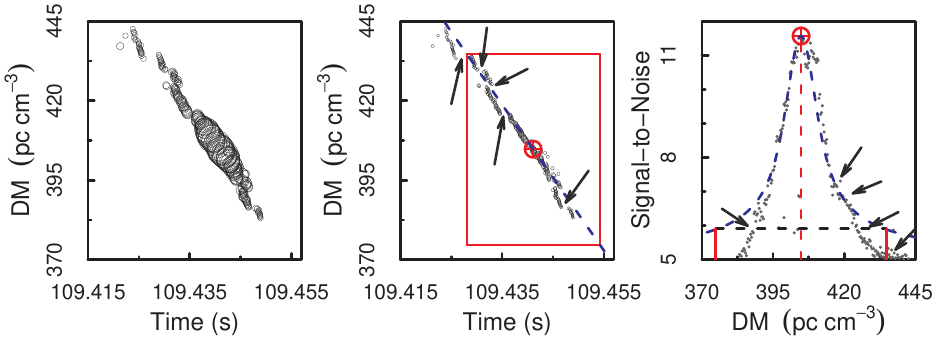}
	\caption{Merging the clusters in a non-clipped pulse. In the left and middle subplots, the single-pulse events are both plotted proportionally to S/N but with different scaling in order to show the distance between single-pulse events more clearly. 
		In the middle subplot, the red circle shows the brightest single-pulse event, and the red solid rectangle shows the expected DM and time span of the pulse.
		In the right subplot, the red circle shows the brightest single-pulse event and the red dashed line shows its DM, and the blue dashed line shows the expected (i.e., not fitted) S/N decline. 
		The black horizontal dashed segment shows the S/N threshold ($\textrm{S}_\textrm{th}$) calculated from equation~(\ref{eq:SNR_thresh}).
		The two red vertical segments show the limit of the DM offset where the S/N decreases below $\textrm{S}_\textrm{th}$.  
		The arrows show where the pulse breaks into several clusters in DBSCAN clustering because a small radius is used.}
	\label{fig:merge_regular_pulse}
\end{figure}

The second reason is the unusually large time drift between two adjacent single-pulse events, which is commonly observed in bright and wide pulses that are made of many single-pulse events. This can cause these pulses to break into multiple clusters during DBSCAN clustering, as shown in Fig.~\ref{fig:merge_regular_pulse}.

Theoretically, using a much larger radius in DBSCAN clustering can solve the problem of ``broken'' SPEGs.
However, one of the requirements set for our approach was that it should be able to separate pulses even when the time difference between them is small. 
Thus, instead of using a much larger radius in DBSCAN clustering that would inevitably group such pulses together, we introduced an extra step of merging the clusters vertically after DBSCAN clustering was completed. 

As described in Section~\ref{sect:spe}, given the peak S/N and width
$W$ of a pulse, we can calculate the expected maximum DM offset that would cause the observed S/N to fall below 5, and hence the expected extent of an astrophysical pulse in DM and time.
Correspondingly, the merging of clusters can be conducted as follows. After DBSCAN clustering, if a dimmer cluster is identified within the expected DM and time span of a brighter cluster, then the dimmer cluster needs to be merged with the brighter one. The merged cluster as a whole is considered as detection of a single pulse.
This suggests that merging should start from the brightest cluster and continue in descending order of maximum S/N. 
However, there are several additional challenges that needed to be addressed, which are listed below.
While the first three challenges mainly relate to non-clipped/regular wide pulses, the last two only relate to clipped pulses.

First of all, for most non-clipped pulses that have a narrow DM extent, merging is unnecessary because extremely large time drift between adjacent single-pulse events is unlikely to be observed among them. Therefore, using $\epsilon =6$ in DBSCAN clustering provides enough robustness for slightly larger time drift and the occasional non-detection of single-pulse events in certain DM channels.

Second, while calculating the expected S/N decline is fast, calculating the fitted S/N decline greatly increases the complexity and run time of our algorithm. 
This is because to find the accurate DM and time span of an astrophysical pulse, parameters including the true peak S/N, the width, and DM all have to be first obtained through NLS regression of equations~(\ref{eq:SNR_DM}) and~(\ref{eq:SNR_DM_erf}) before they can be plugged into the same formulas to calculate the fitted S/N decline. 
On the other hand, for most bright and wide pulses that are detected as many single-pulse events, the difference between the expected S/N decline and the fitted S/N decline is fairly small, as for the example shown Fig.~\ref{fig:typical_pulse}. 
Furthermore, in case of a pulse that forms multiple single-pulse event clusters in DBSCAN clustering (as the one shown in Fig.~\ref{fig:merge_regular_pulse}), theoretically, the SPEG may break at any part, thus it is unlikely that the fitted peak S/N, width, and DM would be significantly more accurate than the observed values. If so, the fitted S/N decline would not be better than its expected counterpart either.
Therefore, having in mind the large number of beams in our full data set, we only calculated the expected S/N decline and used it to find the expected DM and time span of the pulse.

Third, in practice, using 5 as the threshold ($\textrm{S}_\textrm{th}$) for the S/N decline more often than not results in an unnecessarily large DM offset. 
This is mainly because when the S/N decreases closely to the noise level (i.e., S/N = 5) due to large DM offset, the observed S/N begins to decrease much faster than the expected S/N decline, as shown in the right subplot in Fig.~\ref{fig:merge_regular_pulse}. 
This is true even for the fitted S/N decline (see the top right subplot in Fig.~\ref{fig:typical_pulse}).
Such difference can be even larger for brighter pulses. 
Therefore, instead of using a fixed value of 5, 
we derived the following empirical formula (through trial and error) to calculate a dynamic threshold $\textrm{S}_\textrm{th}$: 
\begin{equation}
	\textrm{S}_\textrm{th} = 0.4 \log_2 \textrm{(peak S/N)} + 4.5.
	\label{eq:SNR_thresh}
\end{equation}
This dynamic threshold meets the following criteria:
(i) $\textrm{S}_\textrm{th}$ should be smaller than  $\textrm{peak S/N}$. For any $\textrm{peak S/N} \ge 5.49$, equation~(\ref{eq:SNR_thresh}) always returns $\textrm{S}_\textrm{th}$ that is lower than $\textrm{peak S/N}$. Clusters with $\textrm{peak S/N} < 5.49$ are rarely the starting point of merging because merging is not needed for dim pulses as they are unlikely to break into multiple clusters in DBSCAN clustering, or such clusters are made of single-pulse events that are detected at large DM offsets and therefore should be merged with brighter clusters detected at smaller DM offsets;
(ii) $\textrm{S}_\textrm{th}$ should be greater than 5 in oder to make the calculated DM and time span become smaller, and closer to the actual span;
(iii) $\textrm{S}_\textrm{th}$ should be higher for brighter pulses (i.e., higher peak S/N), thus it should increase as peak S/N increases but at a much slower pace, which explains the introduction of the non-linear log relation. 
The parameters in equation~(\ref{eq:SNR_thresh}) were selected by manually varying the values until they worked well for the majority of non-clipped pulses in our preliminary exploration.

An example of using the threshold obtained from equation~(\ref{eq:SNR_thresh}) to calculate the DM and time span is shown in Fig.~\ref{fig:merge_regular_pulse}.
As can be seen in the middle subplot, the calculated DM and time span is smaller in contrast to using 
S/N = 5
as the threshold. 
In the right subplot in Fig.~\ref{fig:merge_regular_pulse}, a small section of the pulse in the high DM region is not merged and forms a separate SPEG because the observed S/N decline has a longer tail in the high DM region.
Nevertheless, the merged SPEG does include the majority of single-pulse events derived from the pulse, which is good enough for the subsequent steps including peak scoring, feature extraction, and automatic classification. 
Since the S/N of the remaining section monotonically decreases as DM increases, no peak-like shape would be found in this single-pulse event cluster, and therefore it would be filtered in the next step in Section~\ref{sect:scoring}. 
Note that $\textrm{S}_\textrm{th}$ is only used to calculate a more accurate expected DM range in the merging step. It does not serve as a criteria to further filter any cluster.
It is also worth mentioning that, for really bright pulses, the threshold calculated by equation~(\ref{eq:SNR_thresh}) can be a little  conservative (i.e., the calculated DM and time span is smaller than the actual span) that all clusters originating from the same pulse may not be merged by the one containing the brightest single-pulse event. Nevertheless, this is not a problem because the remaining clusters will appear and merge like a clipped pulse, which is described next. Consequently, in SPEGID, we included one extra step of removing duplicate SPEGs (SPEGs having the same brightest single-pulse event) after merging.
\begin{figure}
	\centering\includegraphics[width=\columnwidth]{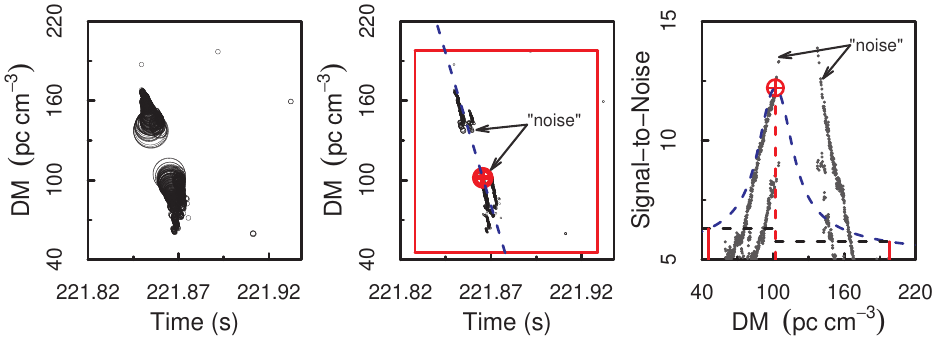}
	\caption{Merging the clipped pulse from Fig.~\ref{fig:clipped_pulses}. The red rectangle in the middle subplot shows the calculated DM and time ranges. 
		Compared with the left subplot, the single-pulse events in the middle subplot are plotted using a smaller scaling to show more details. In the middle and right subplots, the red circle shows the single-pulse event with the maximum S/N among all clusters returned by DBSCAN clustering, from which the calculated DM and time span of the pulse is calculated and shown as the red solid rectangle in the middle subplot.
		Note that the hypothetical peak event is not the brightest event in the right subplot as all the brighter events are excluded from DBSCAN clustering (i.e., there are less than $MinPts$ events in their $\epsilon$ neighborhood, and they don't fall into the $\epsilon$ neighborhood of any other core point) and are classified as ``noise'' (they are not really noise events).	 
		In the right subplot, the blue dashed line is the expected S/N decline calculated from the hypothetical peak.
		The two black horizontal dashed segments show the S/N threshold ($\textrm{S}_\textrm{th}$) calculated from equation~(\ref{eq:SNR_thresh_clipped}).
		The two red vertical segments shows the limit of the DM offset where the S/N decreases below $\textrm{S}_\textrm{th}$.
		By using different $\textrm{S}_\textrm{th}$ obtained on the two sides, the expected DM and time span is closer to the actual span.}
	\label{fig:merge_clipped_pulse}
\end{figure}

The fourth challenge is due to the fact that, in a clipped pulse, the peak S/N and its DM, as well as the pulse width, are all unknown, because the brightest single-pulse event is missing.
As mentioned earlier, calculating the fitted S/N decline would significantly increase the complexity and run time of our algorithm.
Without fitting, the best way to calculate an approximate DM and time span for the clipped pulse is to use the attributes of the brightest single-pulse event of the remaining brightest cluster.
Because this single-pulse event is the closest to the actual clipped peak event, the DM and time span calculated from the former would also be close to that from the latter.
However, the S/N decline calculated from the remaining brightest single-pulse event (using equation~(\ref{eq:SNR_thresh})) tends to decrease too fast towards the side where the pulse was clipped, and too slow towards the other side. This constitutes the last (i.e., fifth) challenge of the merging step.

Finally, when a pulse is clipped, clipping causes
the cluster containing the remaining brightest single-pulse event to end abruptly in the S/N versus DM space. To make the calculated DM and time span closer to the actual span, we adjusted equation~(\ref{eq:SNR_thresh}) to further extend the DM and time extents towards the clipped side and shorten them towards the other side, which is equivalent to shifting the calculated DM and time span towards the clipped side. Specifically, for a clipped pulse, $\textrm{S}_\textrm{th}$ is calculated using the following equation:  
\begin{equation}
	\textrm{S}_\textrm{th} =
	\begin{cases}
		\textrm{S}_\textrm{th} = 0.35 \log_2 \textrm{(peak S/N)}  + 4.5 & \text{on the clipped side,}\\
		\textrm{S}_\textrm{th} = 0.50 \log_2 \textrm{(peak S/N)}  + 4.5 & \text{on the other side.}
		
	\end{cases}
	\label{eq:SNR_thresh_clipped}
\end{equation}
As in equation~(\ref{eq:SNR_thresh}), the parameters in equation~(\ref{eq:SNR_thresh_clipped}) were selected by a trial and error process until they worked well for the majority of clipped pulses in our preliminary exploration. 
Fig.~\ref{fig:merge_clipped_pulse} shows an example of merging a clipped pulse.
As can be seen in the middle subplot, the clusters are successfully merged to form a single SPEG. The right subplot shows that using asymmetric DM and time extents is necessary because the SPEG is no longer symmetric to the hypothetical peak. 

Although in this paper we only demonstrate the effect of merging clusters into SPEGs using data from the PALFA survey, equations~(\ref{eq:SNR_DM}), (\ref{eq:SNR_DM_erf}), (\ref{eq:SNR_thresh}) and (\ref{eq:SNR_thresh_clipped}) are applicable to data from other surveys with no (or minimal) changes of the parameters' values.
Specifically, in equations~(\ref{eq:SNR_DM}) and (\ref{eq:SNR_DM_erf}), the central observing frequency ($\nu$) and the total bandwidth ($\Delta \nu$) are observational setup parameters specific to each survey, whereas the DM offset ($\delta{\textrm{DM}}$), the pulse width ($W$) and observed $\textrm{S/N}$ are all obtained in data processing. 
Therefore, once the observation parameters are given, equations~(\ref{eq:SNR_DM}) and (\ref{eq:SNR_DM_erf}) can be applied to data collected in any survey.
Note, however, that equations~(\ref{eq:SNR_DM}) and (\ref{eq:SNR_DM_erf}) only describe the expected (symmetric, bell-shaped) S/N versus DM curve of astrophysical pulses \citep{Cordes2003}. As this is generally not true for RFI, the calculated DM and time span of RFI may not be as accurate. 
With respect to equations~(\ref{eq:SNR_thresh}) and (\ref{eq:SNR_thresh_clipped}), the criteria we used to formulate these equations should be followed regardless of the observation parameters.
It is important to mention that we tested the equations~(\ref{eq:SNR_DM}), (\ref{eq:SNR_DM_erf}), (\ref{eq:SNR_thresh}) and (\ref{eq:SNR_thresh_clipped}) with no modification (except for the values of $\nu$ and $\Delta \nu$) on data from the Green Bank Telescope (GBT) Drift-scan survey \citep{Karako-Argaman2015}, and they turned out to work equally well.

\begin{figure}
	\centering\includegraphics[width=\columnwidth]{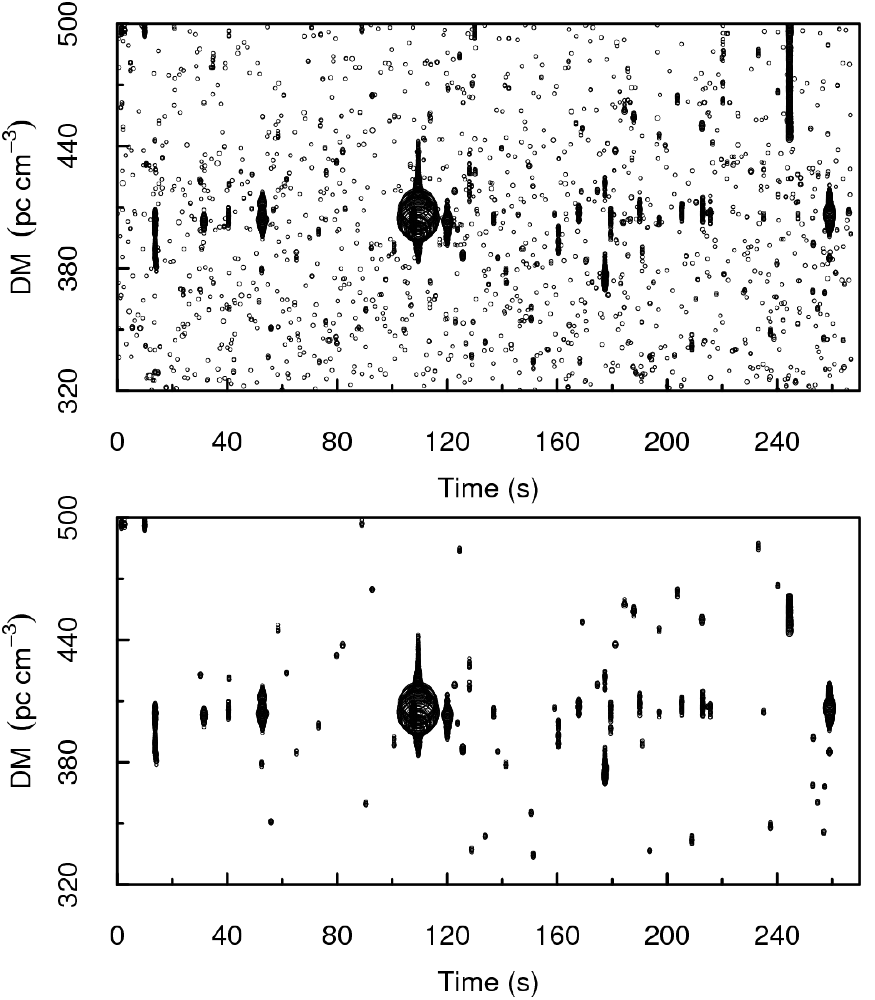}
	\caption{The DM versus time plot for pulsar J1901+02 detected in the PALFA survey. The top subplot shows all single-pulse events within the DM and time range, while the bottom subplot only shows those single-pulse events that belong to SPEGs identified by DBSCAN clustering and merging of clusters. The comparison between the top and bottom subplots illustrates the denoising effect of these two steps.}
	\label{fig:clu_mer_denoise}
\end{figure}

Notice that after merging the clusters vertically, we removed SPEGs that peak at DM $<=$ 2 $\textrm{pc}\ {\textrm{cm}^{-3}} $, which is a general practice to remove RFI \citep{Karako-Argaman2015}.
As illustrated in Figs.~\ref{fig:merge_regular_pulse} and ~\ref{fig:merge_clipped_pulse}, the combination of a small neighborhood radius and merging algorithm successfully identified astrophysical pulses as SPEGs, even if they are broken in the DM versus time space. 
Therefore, in the rest of the paper, we use ``SPEG'' and ``pulse'' interchangeably to refer to both the merged SPEG as a whole and the single-pulse events within. 
Furthermore, as shown in Fig.~\ref{fig:clu_mer_denoise}, our approach effectively eliminates noise that was made up of less than 5 single-pulse events.

\subsubsection{Calculating the peak score for SPEGs}
\label{sect:scoring}
Through DBSCAN clustering of single-pulse events and merging clusters vertically, we obtained a list of SPEGs, among which astrophysical pulses can be identified.
According to equation~(\ref{eq:SNR_DM_erf}), the S/N values of astrophysical pulses peak at the true DM and fall off as the DM offset increases. Searching for this signature within SPEGs can help to discriminate astrophysical pulses against noise and RFI.
In related works, different peak identification algorithms were proposed to identify such trends \citep{Karako-Argaman2015, Devine2016} . 
However, the main challenge of peak identification lies in the variety of the S/N versus DM curves of the pulses (due to varied pulse widths and brightness). 
Our peak scoring algorithm is inspired by the Gaussian shape of the S/N versus DM curve, and two other peak identification algorithms \citep{Karako-Argaman2015, Devine2016}.
A unique characteristic of our algorithm is that it accounts for the fact that different DM spacings are used in different DM ranges, which has been ignored so far. 
This is achieved by making the slope threshold dependent on the DM spacing, which proves to be effective in finding dim and/or wide pulses. 
Furthermore, our peak scoring algorithm is not only capable of identifying the peak in S/N versus DM curves of various shapes, it also returns a score that characterises the shape of a curve.
The peak score conveys additional information about the relative ``steepness'' of the S/N versus DM curve, which is used as one of the features in the automatic classification step (see Section~\ref{sect:feature_extrac}). 

Our peak scoring algorithm calculates the peak score for each SPEG in the following way:

\begin{enumerate}
	\item If the brightest single-pulse event within the SPEG does not have any neighboring event within 5 adjacent DM channels on both sides, the SPEG is then considered as clipped, and it is treated differently from a non-clipped SPEG;
	
	\item 
	Because equations~(\ref{eq:SNR_DM}) and (\ref{eq:SNR_DM_erf}) predict a symmetric S/N versus DM curve for astrophysical pulses,
	we propose two symmetry indices, $\textrm{SI}_\textrm{DM}$ and $\textrm{SI}_\textrm{S/N}$, to characterise the symmetry of the S/N versus DM curve of an SPEG. These two symmetry indices are used as two of the many features that help to distinguish astrophysical SPEGs from non-astrophysical SPEGs. $\textrm{SI}_\textrm{DM}$ and $\textrm{SI}_\textrm{S/N}$ are calculated as follows: 
	\begin{equation}
		\textrm{SI}_\textrm{DM} = \frac{min\textrm{(maxDM -- peakDM, peakDM -- minDM)}}{max\textrm{(maxDM -- peakDM, peakDM -- minDM)}},
		\label{eq:DMSymIndex}
	\end{equation}
	where maxDM is the maximum DM of the SPEG, peakDM is the DM of the brightest single-pulse event and minDM is the minimum DM of the SPEG, and
	\begin{equation}
		\textrm{SI}_\textrm{S/N} = \frac{min\textrm{($\Sigma{S/N_\textrm{left}}$, $\Sigma{S/N_\textrm{right}}$)}}{max\textrm{($\Sigma{S/N_\textrm{left}}$, $\Sigma{S/N_\textrm{right}}$)}},
		\label{eq:S/NSymIndex}
	\end{equation}
	where $\Sigma{S/N_\textrm{left}}$ is the sum of the S/N of all single-pulse events (within the SPEG) to the left of the brightest single-pulse event, and $\Sigma{S/N_\textrm{right}}$ is the corresponding sum of all single-pulse events (within the SPEG) to the right.
	Note that both $\textrm{SI}_\textrm{DM}$ and $\textrm{SI}_\textrm{S/N}$, defined by equations~(\ref{eq:DMSymIndex}) and (\ref{eq:S/NSymIndex}) respectively, have a range between 0 and 1. The higher these two indices are, the more symmetric the SPEG would be.
	In contrast, RFI SPEGs that do not follow the law described in equation~(\ref{eq:SNR_DM}) typically appear to be monotonically decreasing (or increasing) in the S/N versus DM space. Therefore, they usually have symmetry indices close to 0 and can be differentiated from astrophysical SPEGs which are usually more symmetric;
	
	\item For a non-clipped SPEG, $\textrm{SI}_\textrm{DM} > 0.20$ and $\textrm{SI}_\textrm{S/N} > 0.25$ ensure that the brightest single-pulse event of the SPEG is in the middle and the S/N versus DM curve is balanced to a certain degree.
	On the other hand, for a clipped SPEG, only $\textrm{SI}_\textrm{S/N} > 0.10$ is required because clipping often results in less balanced S/N versus DM curve.
	SPEGs (whether clipped or not) that have even lower symmetry indices are considered as RFI and are excluded from further investigation. 
	In practice, these threshold values are so small that only extremely asymmetric SPEGs (that are unlikely to be astrophysical pulses) were eliminated.
	In other words, these threshold values are applicable to surveys with different observational setups;	
	
	\item Inspired by the roughly Gaussian shape of the S/N versus DM curve, we divided a non-clipped SPEG into six sections, with three sections on each side of the brightest single-pulse event. 
	In the case of a clipped SPEG, as it is missing the peak, each side was divided into two sections instead. The same rule also applies to non-clipped SPEGs with less than three single-pulse events on either side;
	
	\item Within each section $i$, we applied weighted least square (WLS) linear regression between S/N and DM and recorded the slope of the regression line as $\textrm{FLS}_\textrm{i}$ (Fit Line Slope). 
	The S/N of each single-pulse event was used as its weight. 
	We chose WLS linear regression over ordinary least squares (OLS) linear regression which was used in RAPID \citep{Devine2016} because the former is more immune to some low S/N noise events that are occasionally included into SPEGs. 
	The comparison between OLS linear regression and WLS linear regression is demonstrated through their application on the clipped pulse shown in Fig.~\ref{fig:wls}. It can be seen that
	the fit lines obtained from WLS linear regression more accurately reflect the actual shape of the S/N versus DM curve. For a clipped pulse, we also recorded the DM of the intersection of the two middle fit lines as the centre DM ($centreDM$), the hypothetical DM of the clipped peak; 
	\begin{figure}
		\centering\includegraphics[width=\columnwidth]{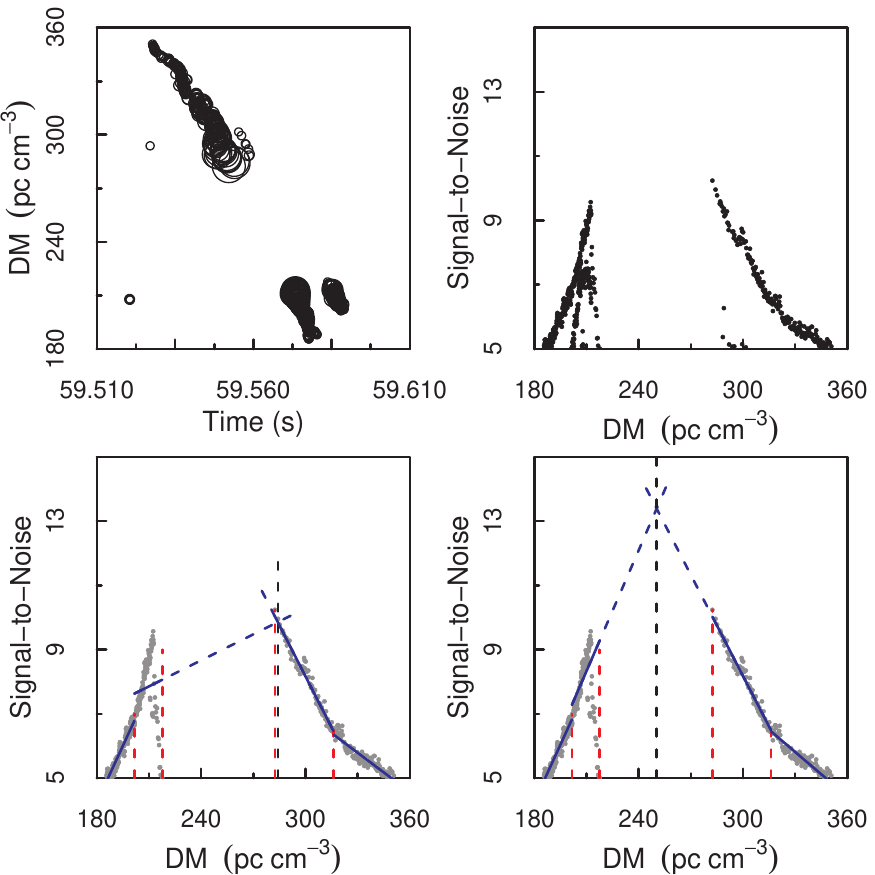}
		\caption{The application of OLS linear regression and WLS linear regression on a clipped pulse. 
			The top left subplot shows the SPEG in the DM versus time space, while the other three plots show the S/N versus DM curve. 
			While the top right subplot shows all single-pulse events within the SPEG, the bottom two plots only keep the brightest event if there is more than one event at the same DM.
			In the bottom left subplot, the fit lines were obtained through OLS liner regression, and the second fit line is strongly influenced by those dim events,
			whereas in the bottom right subplot, their influence is significantly reduced by WLS linear regression.
			The intersection of the middle two fit lines is recored as the centre DM ($centreDM$), and its value is shown by the black dashed line.
			It can be seen that $centreDM$ obtained from WLS linear regression is much closer to the DM of the clipped peak.
		}
		\label{fig:wls}
	\end{figure}
	
	\item For each SPEG, we define a slope threshold $\textrm{SL}_\textrm{th}$ for its fit lines as follows: 
	\begin{equation}
		\textrm{SL}_\textrm{th} = \max(0.01 / {\textrm{DM}_\textrm{spacing},\ 0.01\ \textrm{pc}^{-1}\  {\textrm{cm}^{3}}),}
		\label{eq:slope_thresh}
	\end{equation}
	where $\textrm{DM}_\textrm{spacing}$ is the DM spacing where the S/N peaks and it is defined as: 
	\begin{equation}
		\textrm{DM}_\textrm{spacing} = \textrm{DM}_\textrm{n+1} - \textrm{DM}_\textrm{n},
		\label{eq:DM_spacing}
	\end{equation}
	where $\textrm{DM}_\textrm{n}$ is the DM where the S/N peaks, and $\textrm{DM}_\textrm{n+1}$ is DM of the next trial DM channel.
	Therefore $\textrm{SL}_\textrm{th}$ is DM-spacing-dependent. 
	We made the slope threshold dependent on DM spacing because we believe for a fit line to be considered as steep, its S/N should on average display a minimum change of 0.01 for a DM change of one DM spacing. (Note that we further set a lower bound of 0.01 for $\textrm{SL}_\textrm{th}$ for $\textrm{DM}_\textrm{spacing} > 1 \textrm{pc}^{-1}\  {\textrm{cm}^{3}}$, which is equivalent to requiring a minimum S/N change of 0.01 for a DM change of 1 $\textrm{pc}^{-1}\  {\textrm{cm}^{3}}$.)
	Even smaller change in S/N over one DM step indicates a flat S/N versus DM curve, therefore a score of 0 is assigned to the fit line.
	We selected such a small $\textrm{SL}_\textrm{th}$ because a larger threshold would cause dim and/or very wide pulses to be overlooked. An example of a wide, not very bright pulse is shown in Fig.~\ref{fig:wide_pulse}. It can be seen that the slopes of its fit lines are fairly small.
	
	\begin{figure}
		\centering\includegraphics[width=\columnwidth]{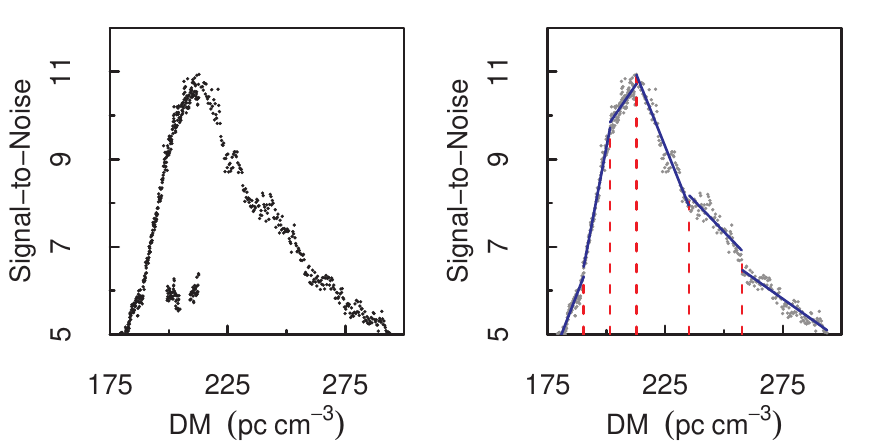}
		\caption{An example of a wide, not very bright pulse. Given the DM spacing at the peak S/N is 0.3 $\textrm{pc}^{-1}\  {\textrm{cm}^{3}}$, $\textrm{SL}_\textrm{th}$ calculated by equations~(\ref{eq:slope_thresh}) and (\ref{eq:DM_spacing}) is 0.03, and the peak score is 6. 
			The slopes of the six fit lines calculated from WLS linear regression are (from left to right): 0.140, 0.283, 0.078, --0.134, --0.056 and --0.038 $\textrm{pc}^{-1}\  {\textrm{cm}^{3}}$.
			Thus using any slope threshold greater than 0.14 will result in a peak score less than 2, and cause this pulse to be overlooked. 
		}
		\label{fig:wide_pulse}
	\end{figure}
	
	\item For each section $i$, we compared the slope of the regression line $\textrm{FLS}_\textrm{i}$ with $\textrm{SL}_\textrm{th}$ of the SPEG, and converted it to a score, namely $\textrm{FLSC}_\textrm{i}$ (Fit Line SCore), using equation~(\ref{eq:fit line-score}):
	\begin{equation}\label{eq:fit line-score}
		\textrm{FLSC}_\textrm{i} =
		\begin{cases}
			-1, & \text{if}\ \textrm{FLS}_\textrm{i} < \textrm{SL}_\textrm{th},\\
			\ \ 1, & \text{if}\ \textrm{FLS}_\textrm{i} > \textrm{SL}_\textrm{th}, \\
			\ \ 0, & \text{otherwise;}
		\end{cases}
	\end{equation}
	\item Finally, we calculated the peak score of the SPEG using equation~(\ref{eq:peak_score}):
	\begin{equation}
		\textrm{Peak Score} = {\sum_\textrm{i=1}^\textrm{n/2}{\textrm{FLSC}}_{\textrm{i}} - \sum_\textrm{i={n/2+1}}^\textrm{n}{\textrm{FLSC}}_{\textrm{i}}},
		\label{eq:peak_score}
	\end{equation}
	where $n$ is the total number of sections that the pulse is divided into (i.e., 6 for a regular pulse and 4 for a clipped pulse). 
	The idea behind equation~(\ref{eq:peak_score}) is that because the S/N versus DM curve of an astrophysical pulse is roughly Gaussian, ideally its slope should be positive on the left half of the curve and negative on its right half. Therefore, if a peak is found in an SPEG, equation~(\ref{eq:peak_score}) would return a positive score;
	otherwise, the returned peak score would be 0 or negative.
\end{enumerate} 

\begin{figure}
	\centering\includegraphics[width=\columnwidth]{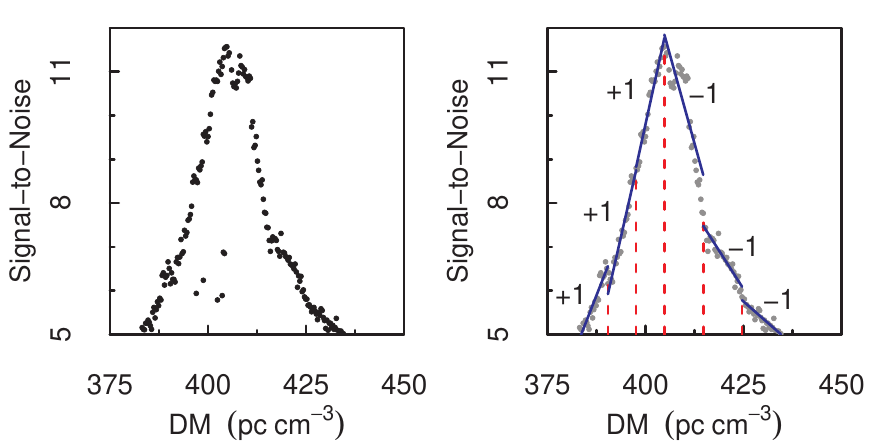}
	\caption{Calculating the peak score of a regular pulse from Fig.~\ref{fig:merge_regular_pulse}. While the left subplot shows all single-pulse events within the SPEG, the right subplot shows only the brightest event if there is more than one event at the same DM. Only these events are used in finding the peak score. The calculated peak score for this pulse is 6.}
	\label{fig:regular_peak_score}
\end{figure}

\begin{figure}
	\centering\includegraphics[width=\columnwidth]{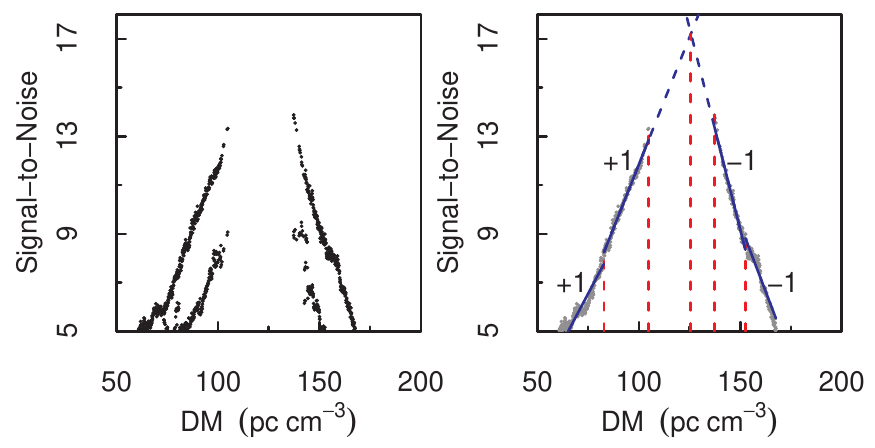}
	\caption{Calculating the peak score of a clipped pulse from Fig.~\ref{fig:merge_clipped_pulse}. 
		While the left subplot shows all events within the SPEG, the right subplot shows only the brightest event if there is more than one event at the same DM. Only these events are used in finding the peak score. The calculated peak score for this pulse is 4.}
	\label{fig:clipped_peak_score}
\end{figure}

\begin{figure}
	\centering\includegraphics[width=\columnwidth]{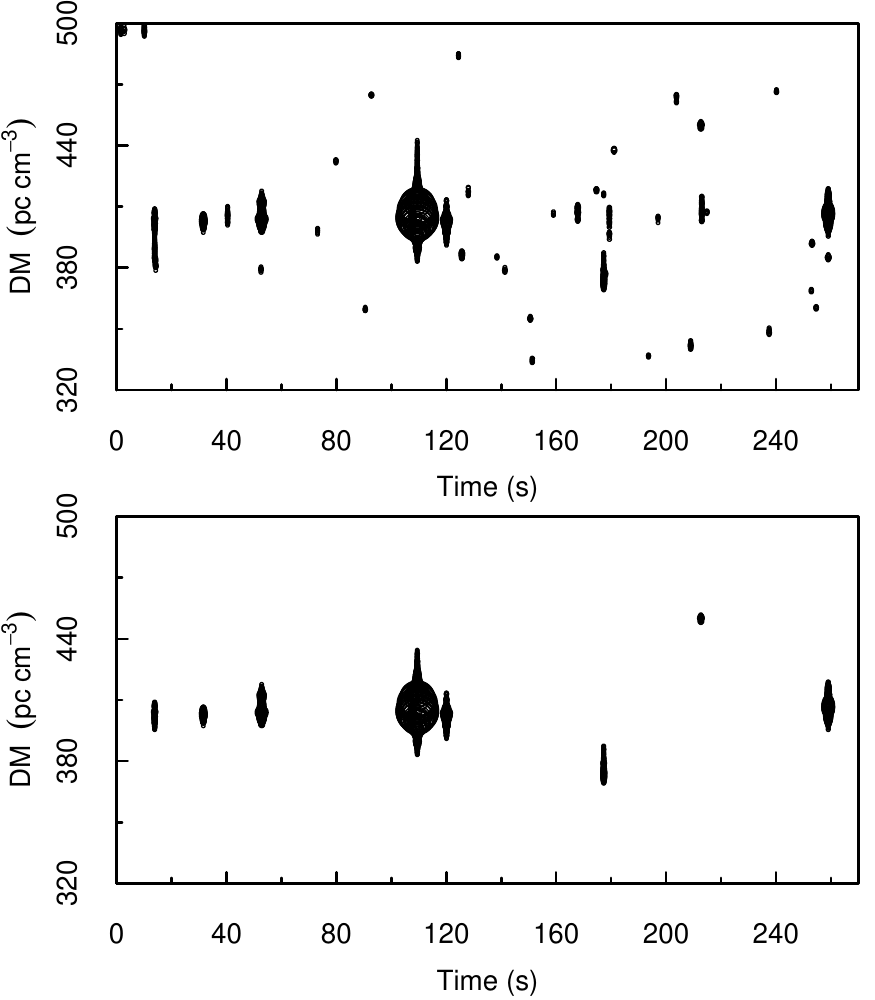}
	\caption{The DM versus time plot for pulsar J1901+02. The top subplot shows all SPEGs with a peak score $\ge$ 2. By filtering SPEGs with a peak score $<$ 2, the number of SPEGs is decreased from 81 (shown in the bottom subplot of Fig.~\ref{fig:clu_mer_denoise}) to 48.
		The bottom subplot only shows those bright SPEGs with a peak score $\ge$ 2 and with a peak S/N $\ge$ 6 (see Section~\ref{sect:DBSCAN} for the definition of bright SPEGs). These SPEGs are used to form new SPEG groups in Section~\ref{sect:grouping}.}
	\label{fig:peak_bright_denoise}
\end{figure}
Typically, a regular bright pulse has a peak score of 6 (as shown in Fig.~\ref{fig:regular_peak_score}), and a clipped pulse has a peak score of 4 (as shown in Fig.~\ref{fig:clipped_peak_score}).
In practice, we found that astrophysical pulses usually have a peak score of at least 2. 
Therefore, by removing all SPEGs with peak score less than 2, we further denoised an observed beam.
An example of such denoising effect is shown in Fig.~\ref{fig:peak_bright_denoise}.
\begin{figure}
	\centering\includegraphics[width=\columnwidth]{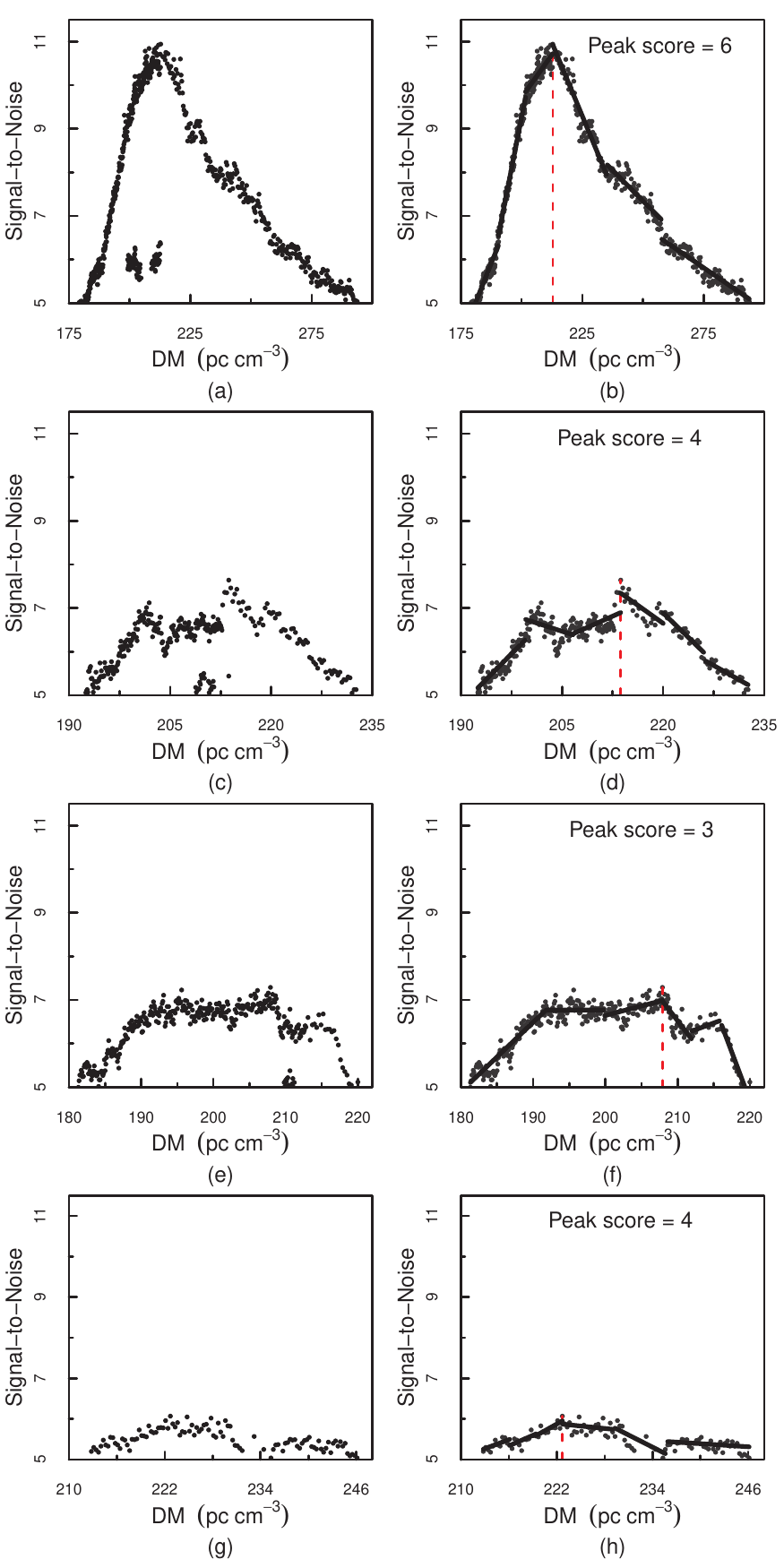}
	\caption{Examples of astrophysical pulses with different S/N versus DM curve shapes (a, c, e, g), and their calculated peak scores (b, d, f, h). These pulses are from the detection of the known pulsar J2003+29 (shown in Fig.~\ref{fig:grouping_pulsar}) in the PALFA survey. Their authenticity is confirmed by searching for the underlying periodicity and comparing it with the period of the pulsar (see Section~\ref{sect:periodicity} for details).
		The S/N versus DM curve in Fig.~\ref{fig:pulse_shapes}(g) looks different than expected, presumably due to the weakness of the pulse.}
	\label{fig:pulse_shapes}
\end{figure}

Compared with RAPID \citep{Devine2016}, we made the following improvements in the peak scoring algorithm of SPEGID: 
\begin{enumerate}
	\item RAPID identifies peaks in the S/N versus DM subplot as DPGs, while in SPEGID the identification of SPEGs is conducted in the DM versus time space (see Sections~\ref{sect:DBSCAN} and~\ref{sect:merging}) and, therefore, separated from calculation of the peak score. 
	\item In SPEGID, we used a lower, DM-spacing-dependent slope threshold. This allowed us to detect wide (Fig.~\ref{fig:pulse_shapes}(a)) and/or dim (Fig.~\ref{fig:pulse_shapes}(c, e, g)) pulses despite variable DM spacing of trial DMs was used in different DM ranges.
	\item In SPEGID, the number of single-pulse events within each section $i$ is not fixed, which is essential for identifying peaks in SPEGs that consist of only a few single-pulse events.
	\item Compared with OLS linear regression which was used in RAPID, using WLS linear regression makes SPEGID more immune to possible low S/N noise events that are occasionally included in SPEGs.
	\item A pulse does not always appear as a single Gaussian peak in the DM versus time space, as those shown in Fig.~\ref{fig:pulse_shapes}(c, g). 
	RAPID may split such a pulse into more than one DPG. 
	In contrast, SPEGID treats this pulse as a whole SPEG (through the merging step, described in Section~\ref{sect:merging}) without breaking it, thus can identify the peak-like shape in the S/N versus DM curve of the pulse.
	\item While RAPID only returns whether a peak was found or not, the peak score adds additional information which is used as a feature in classification stage (see Section~\ref{sect:Stage-2}). 
\end{enumerate}

\subsubsection{Grouping SPEGs horizontally}
\label{sect:grouping}
\begin{figure}
	\centering\includegraphics[width=\columnwidth]{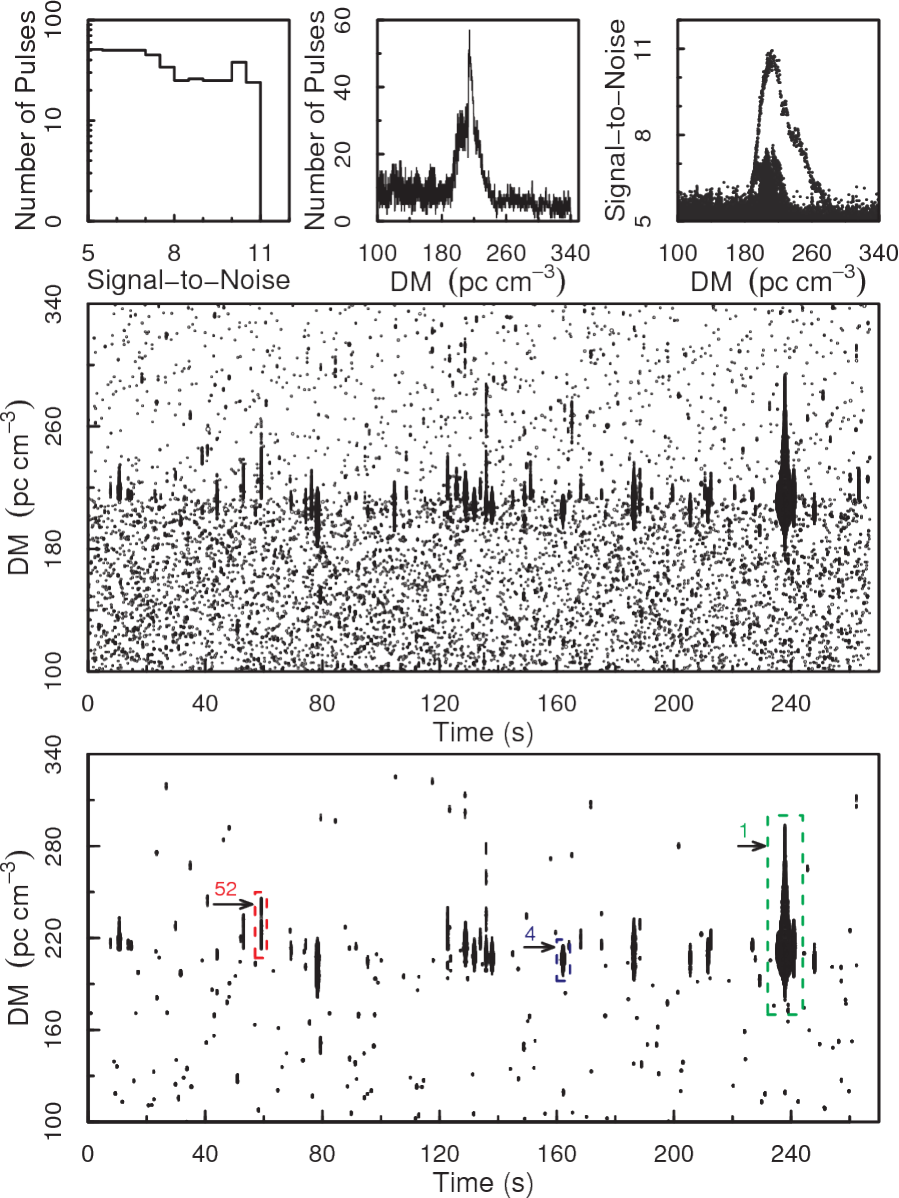}
	\caption{Known pulsar J2003+29 detected on MJD 57039 in the PALFA survey. 
		The three subplots in the top include a histogram of the number of single-pulse events versus S/N, a histogram of the number of single-pulse events versus DM, a scatter plot of S/N versus DM. The middle subplot is a scatter plot of the DM versus time for each single-pulse event where its sise is proportional to its S/N. The bottom subplot shows only SPEGs with peaks identified. 
		The numbers 1, 4 and 52 are the ranks of the SPEGs by their peak S/N in descending order.}
	\label{fig:grouping_pulsar}
\end{figure}

If multiple pulses are detected from the same sky position, a consistent DM will increase the chance that they are astrophysical. Therefore, we developed a method to group SPEGs appearing at a consistent DM. This association between pulses was not considered in any known single-pulse search approach \citep{Karako-Argaman2015, Devine2016}.

\begin{figure}
	\centering\includegraphics[width=\columnwidth]{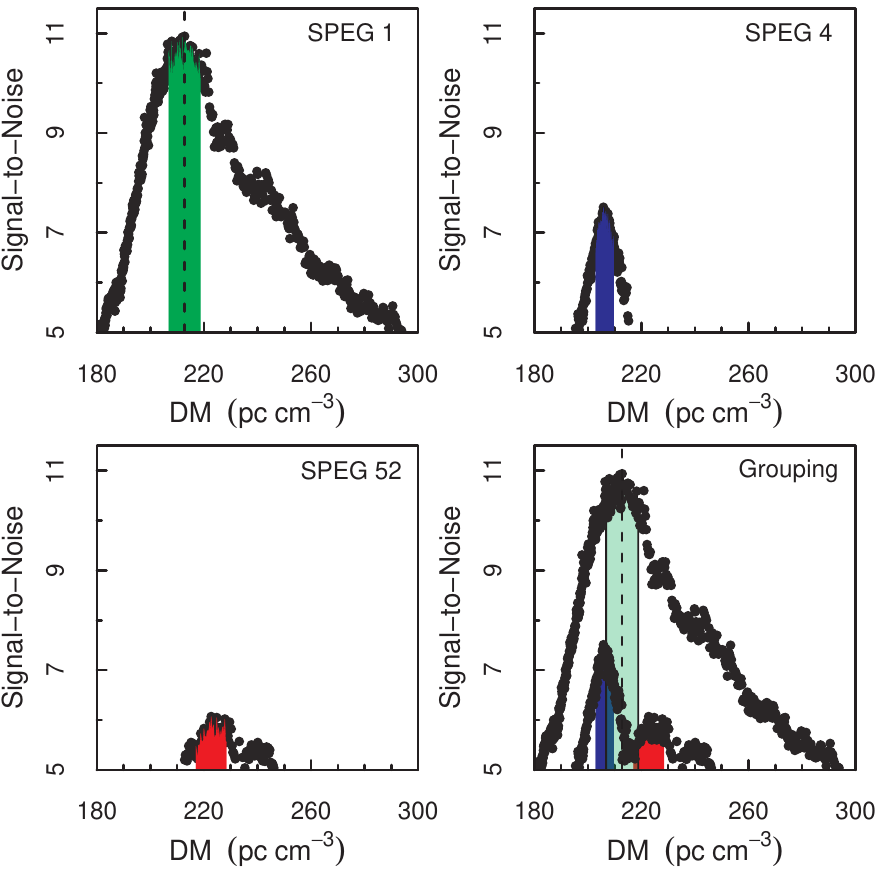}
	\caption{S/N versus DM plots of numbered SPEGs 1, 4 and 52 from Fig~\ref{fig:grouping_pulsar}. The shaded area in each plot shows the central part of the SPEG. In the bottom right plot, the central part of SPEG 1 is plotted as a transparent region (its DM range is represented by the two black solid lines) to show its overlap with that of SPEG 3 more clearly.
		Note that although the DMs of the peak events differ, the central parts of SPEGs 4 and 52 overlap with that of SPEG 1
		and the DM of the peak event of SPEG 1 falls into the full DM ranges of SPEGs 4 and 52. Therefore the three SPEGs are grouped together.}
	\label{fig:grouping_pulses}
\end{figure}

The horizontal grouping of SPEGs is conducted as follows: 
\begin{enumerate}
	\item To account for the fact that the measured DM might vary slightly from pulse to pulse, we define the ``central part'' of an SPEG as the subset of all single-pulse events falling between the highest and the lowest DM events that have an S/N greater than 90\% of the peak S/N. 
	The DM range of this central part is then used to group SPEGs.
	
	\item Starting from the brightest SPEG, we compare the DM range of its central part with that of the rest dimmer SPEGs. 
	If the central DM range of the brightest SPEG overlap with that of any dimmer SPEG, and if the DM of the peak S/N of the brightest SPEG fall within the dimmer SPEG's full DM range, 
	then the two SPEGs are considered as appearing at a consistent DM and be placed into the same group. 
	The first condition ensures that the S/N versus DM curves of the two SPEGs peak at close DMs. The second condition ensures that the periodicity among SPEGs within the same group can be calculated through single-pulse events detected in the DM channel where the S/N of the brightest SPEG peaks (see Section~\ref{sect:periodicity}).
	
	\item This grouping process is continued until the remaining brightest ungrouped SPEG is no longer considered as a ``bright'' SPEG (see Section~\ref{sect:DBSCAN}).
	
	\item Dim SPEGs that are not grouped with any bright SPEG are excluded from further investigation.
\end{enumerate}
Note that the purpose of horizontal grouping is to characterise the association among SPEGs. 
Just like clustering does not guarantee that every noise event will be excluded from all astrophysical pulses,
horizontal grouping alone does not guarantee that all SPEGs from the same source will be placed into the same group, nor that all SPEGs placed in the same group originate from the same source. 
Nevertheless, horizontal grouping of SPEGs provides additional features from SPEG groups. These features, combined with features extracted from individual SPEGs, are likely to improve the automatic classification of SPEGs using supervised machine learning.

\begin{figure}
	\centering\includegraphics[width=\columnwidth]{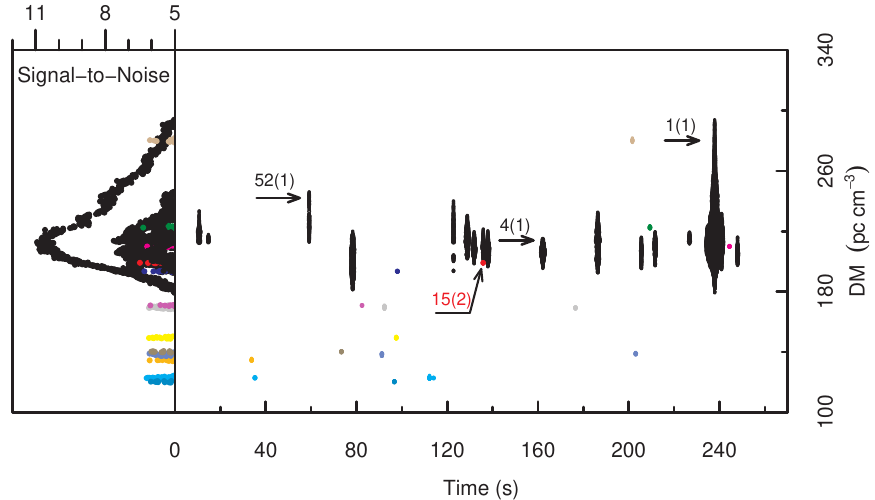}
	\caption{Grouped SPEGs from Fig.~\ref{fig:grouping_pulsar}. Different colors represent different SPEG groups. The number in parentheses is the group number of each SPEG group. 
		For example, SPEG group 1 contains 17 SPEGs (including SPEGs 1, 4, 52, etc.) plotted in black, and SPEG group 2 contains only 1 SPEG (SPEG 15) plotted in red, etc.
		Each group has at least one bright SPEG, with a peak S/N $\ge$ 6.}
	\label{fig:pulse_groups}
\end{figure}

Fig.~\ref{fig:grouping_pulsar} is a diagnostic plot of the known pulsar J2003+29 detected in the PALFA survey. 
Compared with the middle subplot where all single-pulse events are shown, the bottom subplot demonstrates again the denoising effect by only displaying SPEGs with peaks identified (i.e., with peak score $\ge$ 2).
We show how horizontal grouping works on SPEGs in Fig.~\ref{fig:grouping_pulses} using the three numbered SPEGs from Fig.~\ref{fig:grouping_pulsar}.
The resultant SPEG groups are displayed in Fig.~\ref{fig:pulse_groups}, with each SPEG group plotted in a different color.

\subsection{Assessment of the effectiveness of SPEGID}
\label{sect:assessment}
By applying SPEGID on a beam, we removed isolated noise events and identified pulse candidates as SPEGs through DBSCAN clustering and merging of clusters; we also confirmed the peak-like shape in the S/N versus DM curve of SPEGs and filtered out those with peak score less than 2; furthermore, we considered the association between SPEGs by horizontally grouping SPEGs appearing at a consistent DM into the same group. 
These procedures combined together exhibited significant denoising effect in the DM and time space and made astrophysical pulses more noticeable.
As a result, the processed diagnostic plots could be examined with much less effort.
In fact, we manually inspected these plots for all the beams in the benchmark data set (see Section~\ref{sect:benchmark}) and were able to recognise pulsars more easily than in the original diagnostic plots.
However, as mentioned earlier, manual inspection can be both tedious and time consuming because of the large number of diagnostic plots.
Besides, with ever-increasing data in modern surveys, manually inspecting all diagnostic plots becomes impractical. Hence, automatic classification of the identified SPEGs is necessary, 
as done in the second stage of our approach.

\subsection{Automatic classification of SPEGs using supervised machine learning}
\label{sect:Stage-2}

To achieve automatic classification of SPEGs using supervised machine learning, the algorithms are required to acquire the concept of ``pulsar'' from training examples. 
This is an example of concept learning \citep{Mitchell:1997:ML:541177}, which refers to the task of inferring the general definition of a certain concept from training examples that are labelled as members (i.e., pulsars) or nonmembers (i.e., non-pulsars) of the concept. 
Generally speaking, the purpose of concept learning is to find a target function (or target concept) that can correctly map unseen instances to a boolean value based on their features (or attributes). 
Given a set of training examples, learning happens when the learner tries to estimate the target function with a proper hypothesis. 
Hence concept learning can be viewed as the exercise of searching
through the hypothesis space \citep{Mitchell:1997:ML:541177} for one that best approximates the target concept.

In supervised machine learning, as stated in the ``no free lunch'' (NFL) theorem, two supervised machine learning algorithms (i.e., learners) on average exhibit equivalent performance across all possible problems \citep{Wolpert1996, Gomez2016}. 
Recall that concept learning can be seen as searching for the hypothesis that best approximates the target concept; yet, different hypothesis representations (e.g., linear function, decision tree, rule, artificial neural networks, etc) implicitly define different hypothesis spaces. 
Therefore, 
it is necessary to test a wide range of different learners, especially having in mind the lack of research on automatic classification in single-pulse searches.

In the second stage of our approach, to carry out the automatic classification of SPEGs, we first create a fully labelled benchmark data set, which is used to find the best learner and train a classifier with it thereafter. This trained classifier is then used to automatically classify the SPEGs identified from the unlabelled full data set using SPEGID. 
Next, we describe the second stage in detail.

\subsubsection{Feature extraction}
\label{sect:feature_extrac}
Before supervised learning techniques can be applied, we need to extract features from SPEGs, which are listed in Table~\ref{tab:feature_table}.
These features are obtained from three different sources: one feature of the beam, twelve features of individual SPEGs, and five features of the SPEG group. By doing so, our list of features reflect not only the characteristics of individual SPEGs, but also possible association among them. 
Although SPEG groups (from which five features are extracted) include both bright and dim SPEGs, only bright SPEGs (see Section~\ref{sect:DBSCAN}) are further classified.

Note that $\chi^{2}$ of S/N versus DM is usually used to measure how well the shape of the observed pulse in the S/N versus DM space fits the ideal, theoretical shape of a dedispersed pulse given by equation~(\ref{eq:SNR_DM}). However, $\chi^{2}$ was not included in our feature list due to the following two reasons: 
(i) The fitting may take a long time and it does not converge for every SPEG, making it impractical for the large number of SPEGs identified in our data set;
(ii) Previous study showed that $\chi^{2}$ was not a distinguishing feature (i.e., a good predictor) \citep{Devine2016}.

\begin{table}
	\caption{Features extracted for each SPEG and used by machine learning algorithms for classification.}
	\label{tab:feature_table}
	\footnotesize
	\begin{tabular}{c p{0.22\linewidth} p{0.65\linewidth}}
		\hline
		\textbf & {Feature} & \textbf{Description} \\
		\hline
		1 & \textit{ClusterDensity} & Density of clusters in the DM versus time space of this particular beam, defined by equation~(\ref{eq:cluster_density}) \\
		\hline
		2 & \textit{PeakS/N} & Maximum S/N of the SPEG \\
		3 & \textit{PulseWidth} & Width of the brightest single-pulse event within the SPEG, obtained from matched filtering \\
		4 & \textit{CenterDM} & For a regular SPEG, the DM of the peak S/N; for a clipped SPEG, the determined central DM, as described in Section~\ref{sect:scoring} \\ 
		5 & \textit{DMWidth} & SPEG's width in DM (maximum DM of the SPEG minus minimum DM) \\		
		6 & \textit{TimeWidth} & SPEG's width in time (maximum time of the SPEG minus minimum time) \\
		7 & \textit{Clipped} & Boolean value representing whether the SPEG is clipped, described in Section~\ref{sect:scoring} \\
		8 & $\textit{SI}_\textrm{DM}$ & Numerical value measuring the symmetry of the SPEG by DM, defined by equation~(\ref{eq:DMSymIndex}) \\
		9 & $\textit{SI}_\textrm{S/N}$ & Numerical value measuring the symmetry of the SPEG by S/N, defined by equation~(\ref{eq:S/NSymIndex}) \\
		10 & \textit{PeakScore} & Peak score of the SPEG, defined by equation~(\ref{eq:peak_score})  \\
		11 & \textit{SPEGRank} & Rank of the SPEG by \textit{PeakS/N} within the beam \\		
		12 & \textit{WidthRatio} & Ratio of \textit{TimeWidth} over \textit{PulseWidth} \\
		13 & \textit{SizeRatio} & Ratio of number of single-pulse events in the SPEG over number of DM channels in which single-pulse events are detected \\
		\hline
		14 & \textit{GroupMaxS/N} & Peak S/N of brightest SPEG in the group \\
		15 & \textit{GroupRank} & Rank of SPEG group by \textit{GroupMaxS/N} within the beam \\		
		16 & \textit{GroupMedianS/N} & Median of the peak S/N of all SPEGs in the group \\
		17 & \textit{BrightRecurTimes} & Number of bright SPEGs in the group \\
		18 & \textit{RecurTimes} & Total number of SPEGs in the group \\
		\hline
	\end{tabular}
	\label{tab:features}
\end{table}

\subsubsection{Creating the benchmark data set}
\label{sect:benchmark}
In supervised machine learning, the learners have to be trained on a fully labelled training data set before they can be used to classify new (unseen) data.
For that purpose, we carefully created a fully labelled benchmark data set that consists of 90 beams
that included detections of 60
distinct pulsars or RRATs identified in the PALFA survey. These pulsars or RRATs are recorded in either the ATNF pulsar Catalog \citep{ATNF_catalog} or the PALFA New Pulsars List.\footnote{\href{http://www.naic.edu/~palfa/newpulsars/}{http://www.naic.edu/{\raise.17ex\hbox{$\scriptstyle\sim$}}palfa/newpulsars/}}
Our selection criterion was to include as many representative diagnostic plots as possible based on their appearances. 
Because the appearance of a pulsar may vary in different detections, for some pulsars, diagnostic plots of different detections were included.
In addition to the 90 beams with known pulsars, our benchmark includes 900 beams with no known pulsars or RRATs found in them.
If counted by SPEG, our benchmark data set includes 6,386 positively labelled SPEGs and 127,812 non-pulsar SPEGs, all of which were manually labelled.

\subsubsection{Description of the tested machine learning algorithms}
\label{sect:learning_benchmark}
With a goal to select the best performing learner, we test six learners (used in our prior work \citep{Devine2016}) on the benchmark data set: 
JRip,
J48,
PART, 
RandomForest,
SMO (Sequential Minimal Optimization) and MLP (Multilayer perceptron), as listed in Table~\ref{tab:algorithms}.   
All six learners are implemented in Waikato Environment for Knowledge Analysis (\textsc{weka}), which is a popular suite of machine learning software \citep{Weka}.
The six learners represent knowledge in different forms and
exhibit different degrees of interpretability \citep{Witten2011}.

\begin{table}
	\caption{List of machine learning algorithms used for the classification of SPEGs.}
	\centering
	\label{tab:algorithms}
	\footnotesize
	\begin{tabular}{c c}
		\hline
		\textbf{Learner} & \textbf{Type} \\
		\hline
		JRip \citep{Ripper} & Rule \\
		J48 \citep{Quinlan:1993:CPM:583200} & Tree \\
		PART \citep{PART} & Rule + Tree \\
		RandomForest \citep{Breiman2001} & Ensemble Tree \\
		SMO \citep{Platt:1999} & Support Vector Machine \\
		MLP \citep{Murphy:2012:MLP:2380985} & Artificial Neural Network \\
		\hline
	\end{tabular}
	\label{tab:learners}
\end{table}

JRip, which is a \textsc{java}  implementation of the RIPPER (Repeated Incremental Pruning to Produce Error Reduction) algorithm \citep{Ripper}, is a direct rule learner that outputs learned knowledge as rules.
A rule is comprised of two parts, the antecedent (i.e., a series of tests) and the consequent (the class), following a basic ``if antecedent then consequent'' structure.
JRip uses a separate-and-conquer technique to identify rules covering instances from a specific class, separate them out, and continue on the remaining instances. 
To increase the overall accuracy of the learned rule set, the RIPPER algorithm uses incremental reduced-error pruning to generate rules and further revises them using a pruning data set to achieve global optimization.
The learned rules can then be used to classify new instances.
One advantage of a rule learner is that it produces rules that usually can be easily interpreted.

J48, a \textsc{java}  implementation of the C4.5 decision tree algorithm \citep{Quinlan:1993:CPM:583200}, constructs a decision tree using the divide-and-conquer technique. Specifically, attributes are tested sequentially and the instances are split up into subsets based upon the results of each test. 
These tests start from the attribute that results in subsets with the least total information, and continues with the attribute that results in subsets with the second least total information, and so on.
If all instances at a node have the same classification (i.e., information is zero), the development of this part of the tree will stop. 
Like rules, a decision tree can be interpreted by humans.

PART is a hybrid rule-and-tree learner \citep{PART}. Like JRip, PART produces classification rules that are readily interpretable. However, unlike JRip, PART generates each rule by first building a partial decision tree, whose leaf with the largest coverage is made into a rule.
After that, these covered instances are removed and more rules are created for the remaining instances, until all instances are covered.
Obtaining rules from partial decision trees (instead of a fully explored one) avoids the problem of overpruning in a basic rule induction learner and also saves computation time.

RandomForest is an ensemble learning method, which constructs a multitude of decision trees and outputs the average of their classification as the final prediction \citep{Breiman2001}.
The key point is to reduce the correlation among the base learners (decision trees) by constructing each decision tree based on a randomly chosen subset of features extracted from a randomly chosen subset of instances. 
As a result, the variance of prediction would be greatly reduced and thus RandomForest models usually have good predictive accuracy. 
However, using multiple trees generally causes the loss of interpretability.

SMO is a type of support vector machine (SVM) that implements the sequential minimal optimization algorithm for the training of the support vector classifier \citep{Platt:1999}.
SVM first transforms the data into a higher dimensional space using non-linear mapping. 
After that, the algorithm tries to identify the maximum-margin hyperplane that separates different classes in the transformed space and simultaneously has the largest distance from the nearest data points (i.e., support vectors) on both sides. 
This maximum-margin hyperplane corresponds to a non-linear decision boundary in the original data space.
Because SVMs are able to construct complex decision boundaries, they often produce classifiers with high accuracy.
Generally speaking, SVM outputs weights of the features, and the weight of a specific feature indicates its importance in classification.
However, decisions made by SVM are usually difficult to interpret from a human perspective.
Furthermore, training SVM often takes a long time, especially when the data set is large.

MLP is a class of feed forward Artificial Neural Network (ANN) \citep{Murphy:2012:MLP:2380985}.
ANNs identify non-linear decision boundaries of data by including many perceptrons (nodes) that are organized into multiple layers in the network. Each perceptron uses a non-linear activation function.
Such layering allows the building of complex concept and makes MLP well suited for solving complex learning problems.
However, MLP models are usually difficult to interpret and they usually require a large amount of training data.

We use five-fold cross-validation to evaluate these learners on the benchmark data set.
Specifically, the benchmark data set is first divided into five folds using stratified sampling (i.e. each fold contained an equal number of 18 beams with pulsars detected and 180 beams with no pulsar detected).
Every time, different four folds are used to train the learner to build a classifier and the fifth fold is used to test the performance of the classifier.  
The aggregated results of the five folds (reported in Sections~\ref{sect:results_SPEG} and \ref{sect:results_beams}) are then used to select the best learner, based on the F-Measure defined in Section~\ref{sect:metrics}.

\subsubsection{Imbalance consideration}
\label{sect:imbalance}
In our benchmark data set, only 90 out of 990 beams contain astrophysical pulses. 
Furthermore, these 90 beams contain 6,386 positively labelled SPEGs, which account for even smaller proportion of the 134,198 SPEGs identified in the whole benchmark data set.
With a strongly imbalanced distribution of classes like this, many learners can perform poorly regarding the classification of new observations, especially for the minority class (i.e., pulsar class), which is exactly the class of interest.
Consequently, we used Synthetic Minority Oversampling TEchnique (SMOTE) to generate a balanced benchmark data set \citep{chawla2002smote} because our prior work showed that it outperformed other approaches for imbalance treatments \citep{Devine2016}. 
SMOTE increases the size of the minority class by 
creating synthetic instances for each sample from the minority class, and specifically, by adding small random perturbations toward its closest real instances (SPEGs) in the feature space to avoid overfitting. 
When we tested the learners based on the benchmark data set, we only applied SMOTE to SPEGs belonging to the pulsar class in the four folds that were used for training, while the fifth (i.e., testing) fold was used as it was.

\subsubsection{Metrics for evaluation of the classifications}
\label{sect:metrics}
\begin{figure}
	\centering\includegraphics[width=\columnwidth]{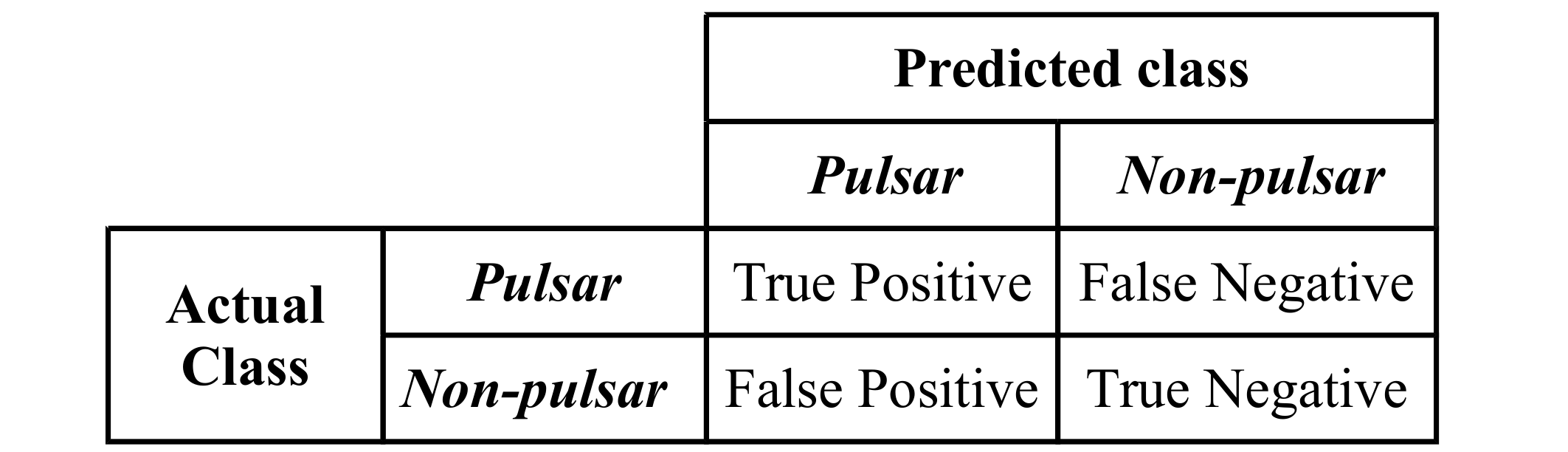}
	\caption{Confusion matrix for SPEG classification.}
	\label{fig:confusion_matrix}
\end{figure}

To measure and compare the effectiveness of learners, we calculated several performance metrics that are derived from the confusion matrix shown in Fig.~\ref{fig:confusion_matrix} \citep{Witten2011}.
The meanings of the four terms in the confusion matrix are as follows:
(i) True Positives (TP): the number of SPEGs that are correctly classified as pulsars; 
(ii) False Positives (FP): the number of SPEGs that are incorrectly classified as pulsars;
(iii) True Negatives (TN): the number of SPEGs that are correctly classified as non-pulsars; 
(iv) False Negatives (FN): the number of SPEGs that are incorrectly classified as non-pulsars.

A good learner produces classifiers with a high number of true positives, i.e., a low number of false negatives. Moreover, a low false positive number is also preferred as it reduces the effort required for manual inspection. 
For each classifier, we calculated the following performance metrics.\\
(i) Recall: the number of positive training instances correctly classified over the total number of positive training instances. Recall is also referred to as the true positive rate (TPR). 
\begin{equation}
	\textrm{Recall} = \frac{\textrm{TP}}{\textrm{TP} + \textrm{FN}}
	\label{eq:TP}
\end{equation}	
(ii) False negative rate (FNR):
the percentage of positive training instances misclassified as belonging to the negative class.
\begin{equation}
	\textrm{FNR} = \frac{\textrm{FN}}{\textrm{FN} + \textrm{TP}}
	\label{eq:FNR}
\end{equation}	
(iii) Precision: the number of positive training instances classified 
correctly over the total number of instances classified as members of the positive class.
\begin{equation}
	\textrm{Precision} = \frac{\textrm{TP}}{\textrm{TP} + \textrm{FP}}
	\label{eq:Precision}
\end{equation}	
(iv) False positive rate (FPR):
the percentage of negative training instances misclassified as belonging to the positive class.
\begin{equation}
	\textrm{FPR} = \frac{\textrm{FP}}{\textrm{FP} + \textrm{TN}}
	\label{eq:FPR}
\end{equation}	
(v) F-Measure (F-M): the harmonic mean of the precision and recall, also referred to as $\textrm{F}_{1}$ Score. 
\begin{equation}
	\textrm{F-M} = 2 \times \frac{\textrm{Precision} \times \textrm{Recall}}{\textrm{Precision} + \textrm{Recall}}
	\label{eq:F_M}
\end{equation}
Because F-Measure is high only if both the precision and recall are high, we used F-Measure to select the best learner. 

\subsubsection{Automatic classification of the unlabelled full data set}
\label{sect:learning_full}
Once the best learner is identified using the benchmark data set, it can be used to classify unlabelled instances.
The automatic classification of unlabelled full data set is   summarised as follows:
(i) We train the best learner on the whole benchmark data set to produce the best classifier (i.e., model);
(ii) We apply SPEGID on the full data set to identify SPEGs and extract features from them;
(iii) We use the best classifier to automatically classify these SPEGs;
(iv) The diagnostic plots of beams containing one or more positively classified SPEGs (by the classifier) are manually inspected to confirm whether these SPEGs candidates are astrophysical.

\subsection{Checking frequency-time signatures for bright SPEGs}
\label{sect:freq-time}
The SPEG candidates automatically classified as belonging to pulsar class (i.e., TP and FP) must be examined manually. 
When multiple bright SPEGs are detected at a consistent DM, their source can be easily confirmed. However, it is not the case for sources with a few dim SPEG candidates being detected.
In practice, we found that if the brightest SPEG candidate has a peak S/N $ \ge$ 8, its astrophysical origin can be verified through inspecting its frequency-time structure. 
\begin{figure}
	\centering\includegraphics[width=\columnwidth]{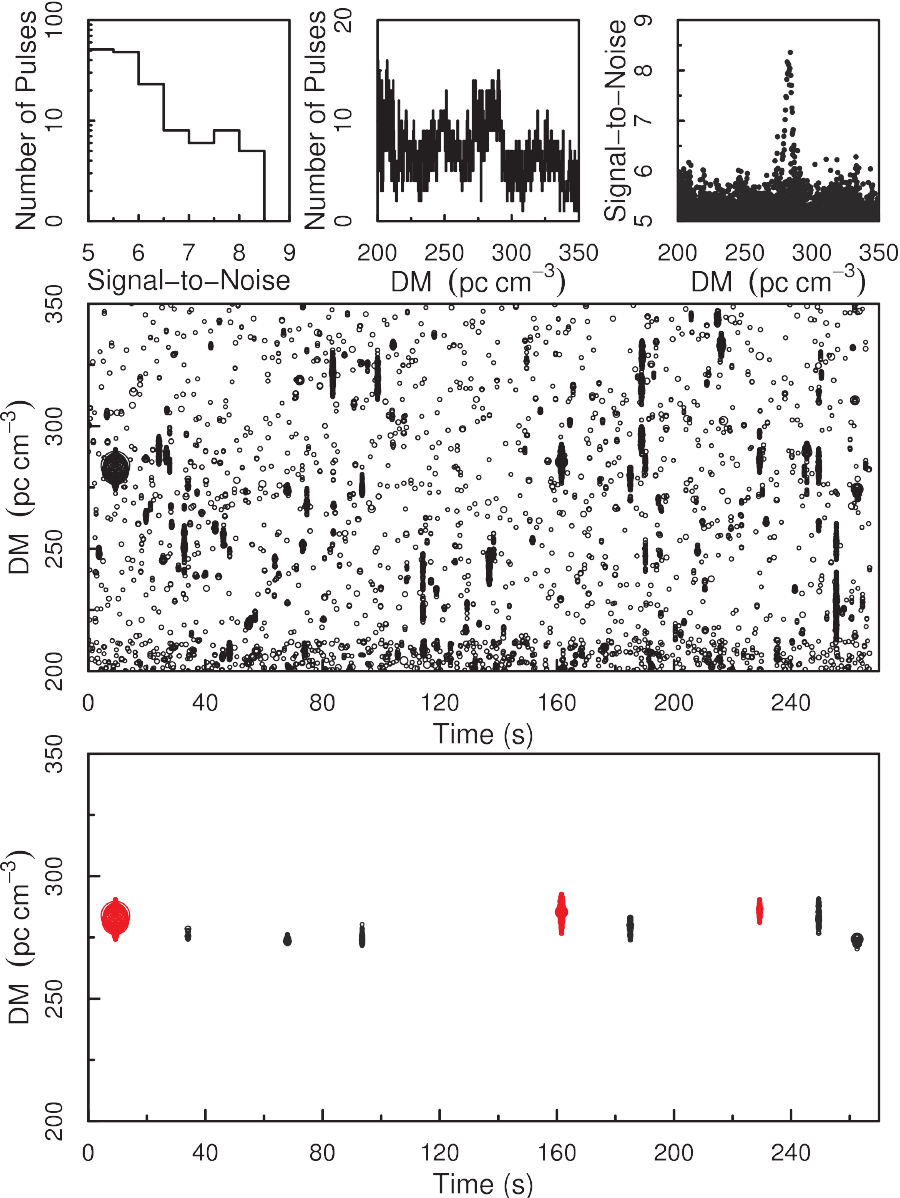}
	\caption{Known pulsar J1910+0728 detected on MJD 56663 in the PALFA survey.
		The bottom subplot shows only SPEGs that have peak score $\ge$ 2 and belong to an SPEG group containing at least one SPEG with peak S/N $\ge$ 6. 
		The red pulses are astrophysical pulses with an underlying period of 0.325 s, which agrees with the period of this pulsar.}
	\label{fig:J1910+0728}
\end{figure}
For example, in Fig.~\ref{fig:J1910+0728} the brightest SPEG at 9.26 s  has a peak S/N of 8.36. By examining its frequency-time signature shown in Fig.~\ref{fig:freq_time}, we confirmed that this pulse was astrophysical, and its attributes agreed with those of the known pulsar J1910+0728. 

\begin{figure}
	\centering\includegraphics[width=\columnwidth]{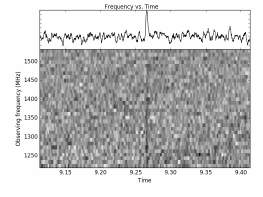}
	\caption{The pulse profile and dynamic spectrum of the pulse detected at 9.26 s in Fig.~\ref{fig:J1910+0728} with maximum S/N = 8.36, dedispersed to DM = 283.9 $\textrm{pc}\ {\textrm{cm}^{-3}} $ and summed to 32 frequency channels across the band. The total time plotted is 300 ms.}
	\label{fig:freq_time}
\end{figure}

\subsection{Searching for periodicity among dim SPEGs}
\label{sect:periodicity}
As mentioned in Section~\ref{sect:grouping}, previous single-pulse search approaches examined each pulse candidate individually without taking into account the association among the candidates \citep{Karako-Argaman2015, Devine2016}. 
For dim SPEGs with S/N $<$ 8, it is difficult to verify whether they are astrophysical pulses or not through the examination of their frequency-time signature. 
For this reason, we developed an approach to search for an underlying periodicity among a group of SPEGs by looking for a common divisor among the gaps of arrival times. The details of this approach are presented in Appendix~\ref{sect:appendix}.
As can be seen in Table~\ref{tab:periodic_probability} (which shows the probability of finding an underlying periodicity among a group of random single-pulse events), when four or more SPEGs are detected at a consistent DM, the probability of them being random events decreases significantly if a periodicity among them is found.

\begin{table}
	\centering
	\caption{The probability of finding an underlying periodicity among random single-pulse events.
		Tolerance level ($\alpha$) reflects how consistent the time differences (between any two neighboring events) are with the predetermined period. Lower $\alpha$ value indicates higher consistency (more details can be found in Appendix~\ref{sect:appendix}).
	}
	
	\label{tab:periodic_probability}
	\begin{tabular}{l| r r r}\hline
		\backslashbox{\textbf{Tolerance level ($\alpha$)}}{\textbf{Number of events}}
		& 3& 4& 5\\\hline
		0.01 &0.955&0.113&0.00161\\\hline
		0.05 &0.992&0.833&0.183\\\hline
	\end{tabular}
\end{table}

From the results listed in in Table~\ref{tab:periodic_probability}, we can make the following conclusions:
\begin{enumerate}
	\item A periodicity will almost certainly be found among three single-pulse events;
	\item When a highly accurate period is found among four or more single-pulse events, or a less accurate period is found among five or more single-pulse events, these events are more likely to be astrophysical signals than random noise.
\end{enumerate}

\begin{figure}
	\centering\includegraphics[width=\columnwidth]{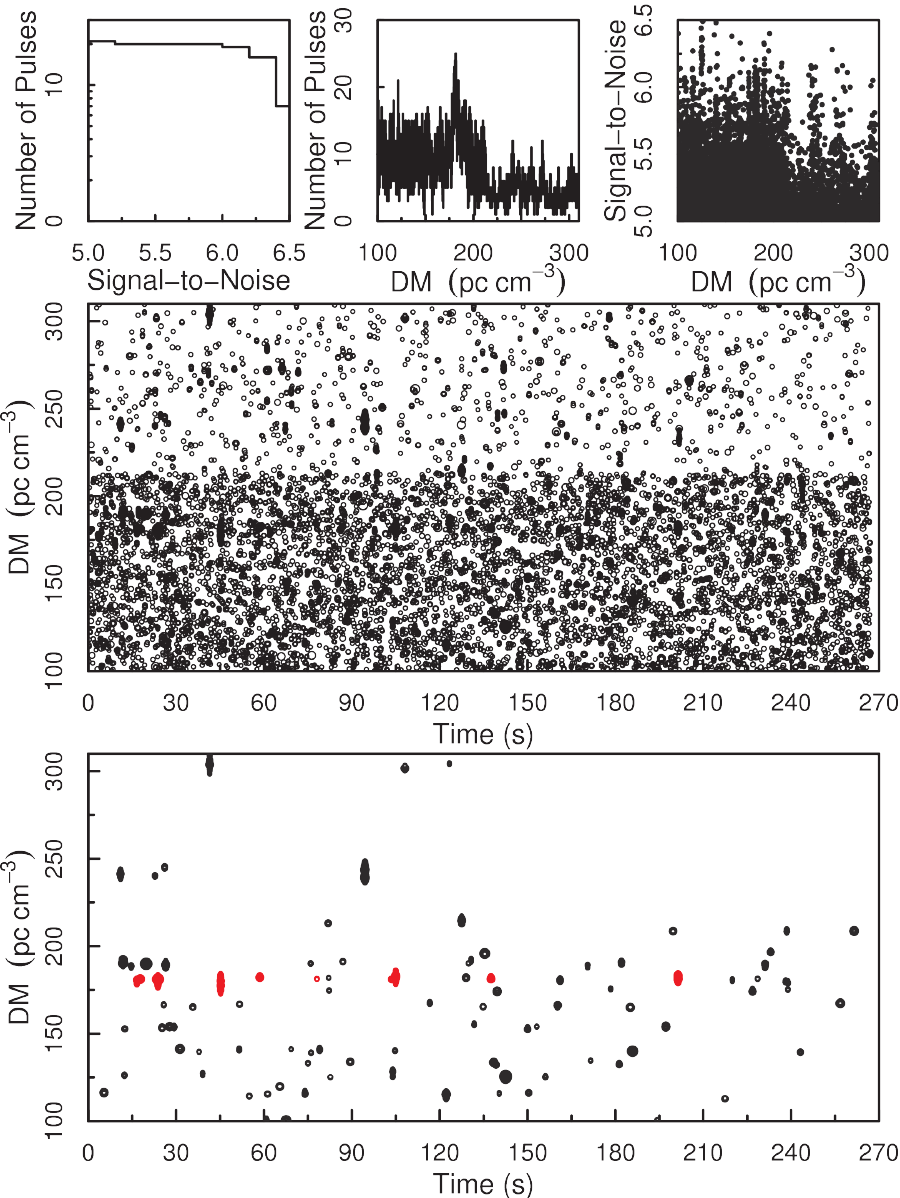}
	\caption{Detection of known pulsar J1935+2025 on MJD 56431 in the PALFA survey. 
		The top three and the middle subplots make up the standard diagnostic plot, in which all single-pulse events within the DM range are plotted. 
		It can be seen that it is difficult, if not impossible, to identify the astrophysical pulses (plotted in red in the bottom subplot) when this plot is manually inspected.
		In contrast, the bottom subplot only shows SPEGs that have peak score $\ge$ 2 and belong to SPEG groups containing at least one SPEG with peak S/N $\ge$ 6, in which the astrophysical pulses can be more easily identified.
		These pulses are plotted in red and they have an underlying period of 0.080133 s, which is the same as the period of pulsar J1935+2025.
		Although dim SPEGs are ignored in the classification step, 
		when periodicity is calculated, both bright and dim SPEGs within the group are considered. 
		Including dim SPEGs is necessary to avoid missing a series of pulses that are mostly dim. 
		Moreover, because the DM spacing increases from 0.1 to 0.3 $\textrm{pc}\ {\textrm{cm}^{-3}}$ at DM = 213.1 $\textrm{pc}\ {\textrm{cm}^{-3}}$, it can be seen that the overall density of single-pulse events in the DM versus time space decreases significantly in the upper half of the middle subplot.	
	}
	\label{fig:periocic_pulses}
\end{figure}
An example of a pulsar with only a series of fairly dim pulses being detected is shown in Fig.~\ref{fig:periocic_pulses}. 
Our approach identified ten SPEGs in this diagnostic plot with an underlying periodicity (at $\alpha = 0.02$) at the DM of 181.1 $\textrm{pc}\ {\textrm{cm}^{-3}} $. 
The brightest SPEG only has a peak S/N of 6.58, which makes checking its frequency-time signature to confirm its astrophysical nature difficult (see Section~\ref{sect:freq-time}). 
However, we found a period of 0.080133 s among these ten SPEGs, and the position of the beam is close to the known pulsar J1935+2025 at the DM of 182 $\textrm{pc}\ {\textrm{cm}^{-3}} $ with a period of 0.080118 s. 
Therefore, we conclude that these ten pulses were from the pulsar J1935+2025. 

Note that we included recurrence times (instead of whether the SPEG belonged to a periodic SPEG group or not) as a feature for classification (see Table~\ref{tab:features}). This is because the probability of finding periodicity among three or four single-pulse events is high, and when the number of SPEGs within an SPEG group is large, calculating periodicity becomes time consuming. 
Hence we only searched for periodicity among SPEG groups that contained one or more positively classified SPEGs.
If an underlying periodicity was identified, we then visually inspected the SPEGs and recorded the most promising candidates for future follow-up observation. 
In contrast, SPEG groups that consisted of mostly dim SPEGs with no underlying periodicity found among them, were classified as RFI.

\section{Results}
\label{sect:results}
In this section, we first evaluate the performance of the six learners on the benchmark data set, using the metrics described in Section~\ref{sect:metrics}, with the goal of finding the best learner.
Then we present the classification results of running the model (which is obtained by training the best learner on the whole benchmark data set) on the unlabelled full data set consisting of 47,042 independent beams that are observations of 33,536 unique positions from the PALFA survey. 

Note that we measured the learners' performance both by SPEG/pulse and by beam. 
Typically, the confusion matrix (and the corresponding performance metrics) based on SPEGs is likely to differ from the confusion matrix based on beams. This is because multiple pulses are often detected from the same pulsar, and these pulses would be identified by SPEGID as separate SPEGs. 
Because we manually inspected the beams that contained one or more positively classified SPEGs (by our trained model), as long as one of these SPEGs was correctly classified, we could find the pulsar in the beam.
In other words, the misclassification of one or more SPEGs during the automatic classification process does not necessarily result in missing the pulsar automatically. 
On the other hand, misclassification of some SPEGs is inevitable in both automatic classification and manual inspection.
Therefore, we selected the best learner based on the results by beam rather than by SPEG, as our goal was to find as many pulsars (rather than SPEGs) as possible.

\subsection{Results based on the benchmark data set by SPEG}
\label{sect:results_SPEG}
\begin{table}
	\centering
	\footnotesize
	\caption{The results based on the benchmark data set by SPEG.}
	\label{tab:benchmark_cluster_stat}
	\begin{tabular}{l l r r r r }
		\hline		
		\textbf{Classifier}  & \textbf{Treatment} & \textbf{TP} & \textbf{FN} & \textbf{FP} & \textbf{TN} \\
		\hline
		$JRIP$&none&5787&599&216&127596\\
		$J48$&none&5855&531&659&127153\\
		$PART$&none&6154&232&221&127591\\
		$RandomForest$&none&6029&357&54&127758\\
		$SMO$&none&3585&2801&327&127485\\
		$MLP$&none&5614&772&140&127672\\
		
		\\
		$JRIP$&SMOTE&6136&250&425&127387\\
		$J48$&SMOTE&6107&279&561&27251\\
		$PART$&SMOTE&6142&244&399&127413\\
		$RandomForest$&SMOTE&6191&195&212&127600\\
		$SMO$&SMOTE&5842&544&4658&123154\\
		$MLP$&SMOTE&6057&329&772&127040\\
		
		\hline
	\end{tabular}
\end{table}

\begin{table}
	\centering
	\footnotesize
	\caption{The performance metrics based on the benchmark data set by SPEG.}
	\label{tab:benchmark_cluster_metric}
	\setlength\tabcolsep{3pt}
	\begin{tabular}{l r r r r r r r}
		\hline
		\textbf{Classifier} & \textbf{Treatment }& \textbf{Recall} &  \textbf{FNR} &  \textbf{Precision} & \textbf{FPR} & \textbf{F-M} \\
		\hline
		$JRIP$&none&0.906&0.094&0.964&0.002&0.934\\
		$J48$&none&0.917&0.083&0.899&0.005&0.908\\
		$PART$&none&0.964&0.036&0.965&0.002&0.965\\
		$RandomForest$&none&0.944&0.056&0.991&0.000&0.967\\
		$SMO$&none&0.561&0.439&0.916&0.003&0.696\\
		$MLP$&none&0.879&0.121&0.976&0.001&0.925\\
		
		\\
		$JRIP$&SMOTE&0.961&0.039&0.935&0.003&0.948\\
		$J48$&SMOTE&0.956&0.044&0.916&0.020&0.936\\
		$PART$&SMOTE&0.962&0.038&0.939&0.003&0.950\\
		$RandomForest$&SMOTE&0.969&0.031&0.967&0.002&0.968\\
		$SMO$&SMOTE&0.915&0.085&0.556&0.036&0.692\\
		$MLP$&SMOTE&0.948&0.052&0.887&0.006&0.917\\
		
		\hline
	\end{tabular}
\end{table}

Table~\ref{tab:benchmark_cluster_stat} shows the confusion matrix of the six learners tested on the benchmark data set, and Table~\ref{tab:benchmark_cluster_metric} shows the corresponding performance metrics. From these two tables, we can make the following observations:
\begin{enumerate}
	\item Without imbalance treatment, five of the six learners performed reasonably well. SMO was the worst classifier that missed a large portion of SPEGs in the pulsar class;
	\item With SMOTE treatment, the Recall of five of the six learners increased, while that of PART showed little change;
	\item When measured by F-Measure, RandomForest with SMOTE treatment proved to be the best learner overall. This agrees with our previous study \citep{Devine2016}.
\end{enumerate}

\subsection{Results based on the benchmark data set by beam}
\label{sect:results_beams}
As mentioned before, the confusion matrix (and the corresponding performance metrics) based on beams may differ from that based on SPEGs because multiple pulses are often detected from the same pulsar.
In order to investigate potential differences, we 
show the confusion matrix and the performance metrics of the same experiments based on beams in Tables~\ref{tab:benchmark_pulsar_stat} and~\ref{tab:benchmark_pulsar_metric}, respectively. 
\begin{table}
	\centering
	\footnotesize
	\caption{The results based on the benchmark data set by beam.}
	\label{tab:benchmark_pulsar_stat}
	\begin{tabular}{l l r r r r }
		\hline			
		\textbf{Classifier}  & \textbf{Treatment} & \textbf{TP} & \textbf{FN} & \textbf{FP} & \textbf{TN} \\
		\hline
		$JRIP$&none&83&7&22&878\\
		$J48$&none&82&8&32&868\\
		$PART$&none&84&6&38&862\\
		$RandomForest$&none&81&9&3&897\\
		$SMO$&none&41&49&27&873\\
		$MLP$&none&81&9&29&871\\
		
		\\
		$JRIP$&SMOTE&84&6&53&847\\
		$J48$&SMOTE&86&4&98&802\\
		$PART$&SMOTE&88&2&84&816\\
		$RandomForest$&SMOTE&86&4&18&882\\
		$SMO$&SMOTE&88&2&687&213\\	
		$MLP$&SMOTE&87&3&193&707\\
		
		\hline
	\end{tabular}
\end{table}
\begin{table}
	\centering
	\footnotesize
	\caption{The performance metrics based on the benchmark data set by beam.}
	\label{tab:benchmark_pulsar_metric}
	\setlength\tabcolsep{3pt}
	\begin{tabular}{l r r r r r r r}
		\hline
		\textbf{Classifier} & \textbf{Treatment }& \textbf{Recall} &  \textbf{FNR} &  \textbf{Precision} & \textbf{FPR} & \textbf{F-M} \\
		\hline
		$JRIP$&none&0.922&0.078&0.790&0.024&0.851\\
		$J48$&none&0.911&0.089&0.719&0.036&0.804\\
		$PART$&none&0.933&0.067&0.689&0.042&0.792\\
		$RandomForest$&none&0.900&0.100&0.964&0.003&0.931\\
		$SMO$&none&0.456&0.544&0.603&0.030&0.519\\
		$MLP$&none&0.900&0.100&0.736&0.032&0.810\\
		
		\\
		$JRIP$&SMOTE&0.933&0.067&0.613&0.059&0.740\\
		$J48$&SMOTE&0.956&0.044&0.467&0.109&0.628\\
		$PART$&SMOTE&0.978&0.022&0.512&0.093&0.672\\
		$RandomForest$&SMOTE&0.956&0.044&0.827&0.020&0.887\\
		$SMO$&SMOTE&0.978&0.022&0.114&0.763&0.203\\
		$MLP$&SMOTE&0.967&0.033&0.311&0.214&0.470\\
		
		\hline
	\end{tabular}
\end{table}
Compared with the results by SPEGs, similar trends can be seen here. However, there are several different trends worth mentioning, which are described next. 

When measured by beam, the six learners showed comparable Recall (hence the FNR) to the results based on SPEGs. However, the FPR of the six learners based on beams increased significantly. Such increase in FPR indicates that SPEGs that were incorrectly classified as pulsars were distributed among many beams.

With SMOTE treatment, the FNR of all six learners decreased to below 5\%, but the FPR increased more significantly. Consequently, the F-Measure of all six learners with SMOTE treatment decreased. We still used SMOTE because, in case of pulsar classification, the learner with a lower FNR is preferred (when they have comparable FPR).

Among the six learners with SMOTE treatment, RandomForest showed the highest F-Measure. Since all six learners had over 93\% recall when SMOTE treatment was applied, the difference in F-Measure mainly originated from the difference in precision.  
Notably, RandomForest had a 2.2\% lower recall compared to PART and SMO, but a significantly higher precision (i.e., 82.7\% compared to 51.2\% and 11.4\%). This means that with slightly lower recall (i.e., slightly higher FNR), among the positively classified instances RandomForest produced significantly more true positive than false positive instances.  
RandomForest showed the highest precision (82.7\%) and the lowest FPR (2.0\%) most likely because it mitigated the problem of overfitting by building multiple decorrelated learning trees and outputting the average of their classification as the final prediction.
In contrast, overfitting is commonly seen in decision tree methods and artificial neural networks as they tend to fit the training data very well \citep{Pham2008}.
Using multiple trees makes the RandomForest classifier lose the interpretability properties (compared to J48 and PART for example), but this is not a problem since we are more concerned with the predictive power of the classifiers.

If 9.3\% FPR was considered acceptable, one could argue that PART is the best classifier. As for SMO, the extremely high FPR (76.3\% when measured by beam) is likely due to the fact that some dim RFI and/or noise look very similar to astrophysical pulses. Consequently, not only it is hard to differentiate them through manual inspection, the algorithm could also have difficulties identifying effective support vectors that decide the maximum-margin hyperplane that separates the instances in the feature space.

Based on the highest F-Measure (i.e., both recall and precision being high) and lowest FPR, we selected RandomForest with SMOTE as the best learner to be applied on the unlabelled full data set. 
By examining the four pulsars that were misclassified by RandomForest with SMOTE treatment, we found that three of them were dim pulsars with less than five pulses detected, while the fourth pulsar was detected at the lowest DM (8.0 $\textrm{pc}\ {\textrm{cm}^{-3}}$) among all known pulsars in the benchmark. In other words, these pulsars were missed because they are different from the other pulsars included in the benchmark data set. 

\subsection{Results based on the full data set}
%
Before the trained model can be used to classify the unlabelled full data set, we first processed the full data set using SPEGID described in Section~\ref{sect:approach} to identify SPEGs and extract features from them. 
Then, we used the best model (obtained by training the selected best learner -- RandomForest with SMOTE treatment -- on the whole benchmark data set) to classify the SPEGs identified in the unlabelled full data set.  
The results are listed in Table~\ref{tab:full_cluster_stat}.

\begin{table}
	\centering
	\caption{The results based on the full data set by SPEG.}
	\label{tab:full_cluster_stat}
	\begin{tabular}{l l r r }
		\hline			
		\textbf{Classifier}  & \textbf{Treatment } & \textbf{Total}& \textbf{Positive} \\
		\hline
		$RandomForest$ & SMOTE & 7,255,112 & 24,426\\		
		\hline
	\end{tabular}
\end{table}

From Table~\ref{tab:full_cluster_stat} it can be seen that 
in the SPEG identification stage, 
from the 47,042 beams in the full data set, our proposed approach identified 7,255,112 SPEGs with peak score $\ge$ 2. Out of these, 24,426 SPEGs (from 1,299 beams) were classified as pulsars.

Since it was impractical and unnecessary to examine all 24,426 positively classified SPEGs, we manually examined the diagnostic plots of the 1,299 beams from which these 24,426 SPEGs were identified and only present the statistics based on beams here. 
(Because the same pulsar can be detected in multiple beams, unless we point it out specifically, here we use ``pulsar'' to refer to the detection of a pulsar within a beam.) 
Through quick-look manual inspection, the 1299 beams that contained positively classified SPEG candidates were distributed across four categories as follows: 
90 known pulsars (60 distinct) included in the benchmark data set were correctly classified (CKs);
81 additional known pulsars not included in the benchmark data set were correctly classified (AKs); 
1031 false positives, i.e., non-pulsars incorrectly classified as pulsars (FPs); and 
97 new candidates of interest (NCs).

\begin{figure}
	\centering\includegraphics[width=\columnwidth]{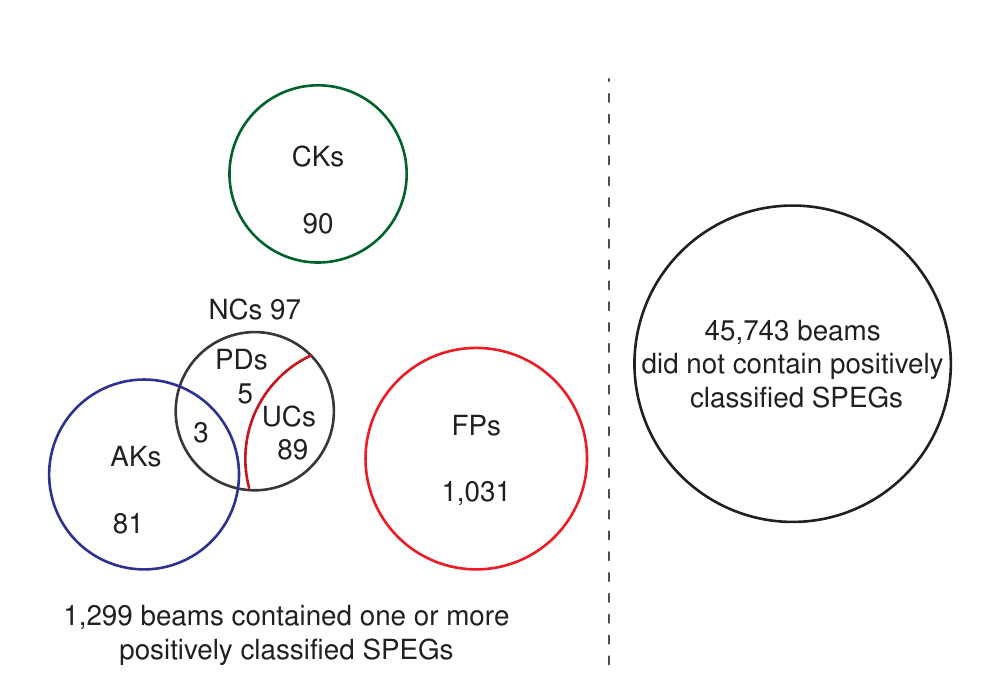}
	\caption{
		A Venn diagram showing the results of automatic classification and manual inspection of beams that contained positively classified SPEG(s). 
		Among these 47,042 beams, only 1,299 beams contained at least one positively classified SPEG. 
		Through quick-look manual inspection of these 1,299 beams, we found all 90
		known (60 distinct) pulsars included in the benchmark data set correctly classified (CKs), 
		84 additional known pulsars not included in the benchmark data set (AKs),
		1031 false positives (FPs), i.e., non-pulsars incorrectly classified as pulsars), 
		89 unlikely candidates (UCs), and 
		5 possible discoveries (PDs).}
	\label{fig:results_Venn}
\end{figure}

\begin{figure}
	\centering
	\includegraphics[width=\columnwidth]{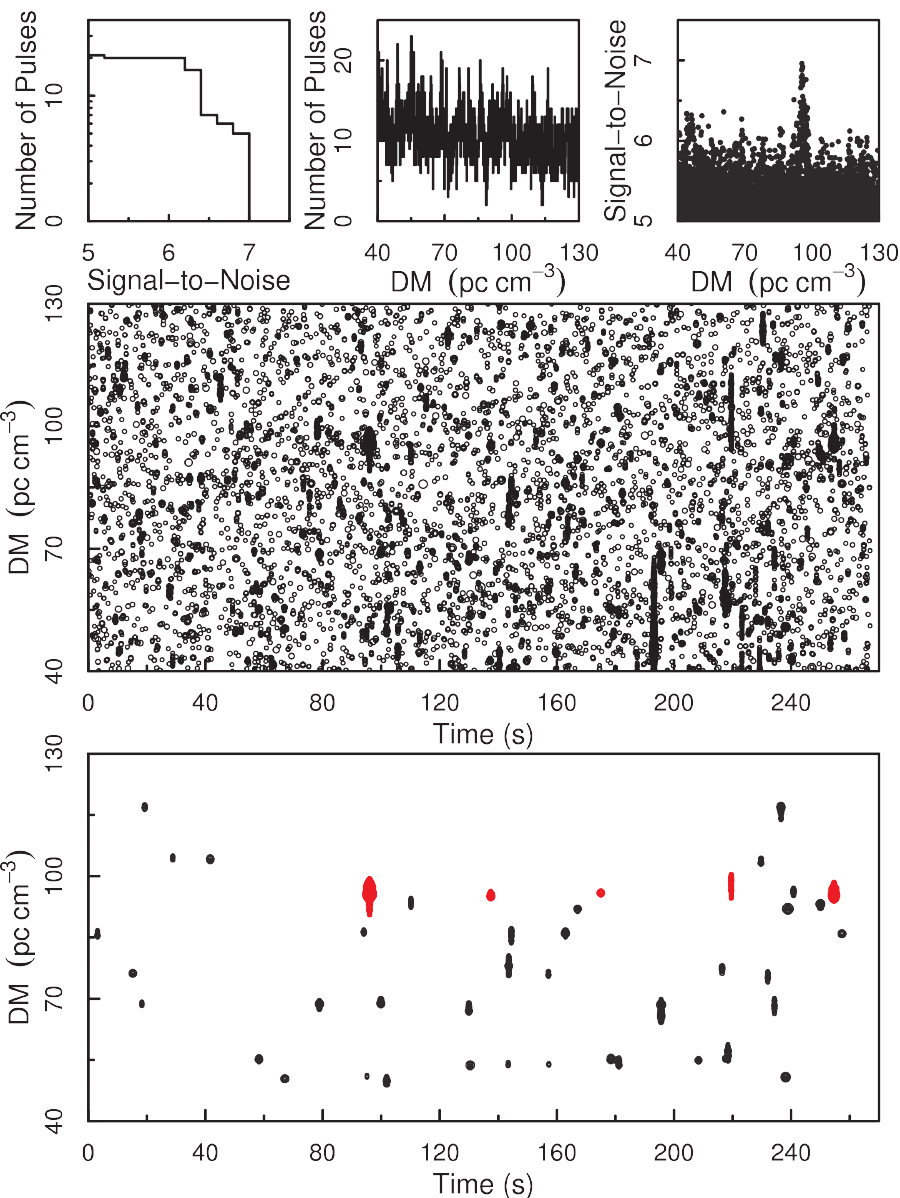}
	\caption{One possible discovery. The five SPEGs plotted in red in the bottom subplot are ``periodic'' at the tolerance level of 0.03. The probability of  finding periodicity among five random events at such tolerance is 0.043. 
		Nevertheless, it is worth pointing out that the candidate SPEGs are made up of many single-pulse events with a peak-like shape found in their S/N versus DM curve, therefore, the probability of them being purely random should be significantly lower.
	}
	\label{fig:possible_discovery}
\end{figure}

The list of NCs mainly consisted of SPEG groups that were made up of a few fairly dim SPEGs (see Figs.~\ref{fig:periocic_pulses} and \ref{fig:possible_discovery}),
which is not surprising as bright pulsars are much easier to detect and consequently it is more likely that they have already been discovered.
With the aid of techniques described in Sections \ref{sect:freq-time} and \ref{sect:periodicity}, out of the 97 NCs we confirmed 3 additional AKs (making the total number of AKs increase to 84). We also recorded 5 promising candidates as Possible Discoveries (PDs) and classified the rest 89 beams as Unlikely Candidates (UCs). 
The results are presented in a Venn diagram shown in Fig.~\ref{fig:results_Venn}.

From these results, the following conclusions can be made:
\begin{enumerate}
	\item Compared to the total number of 47,042 beams in our data set, our approach reduced the number of beams that required manual inspection to 1,299 (i.e., by over 97\%);
	\item Our classifier successfully found all 90 beams containing (60 distinct) known pulsars in the benchmark data set;
	\item Our classifier was able to find 84 additional beams containing 49 distinct known pulsars. Among these 49 pulsars, 17 were already included in the benchmark data set. In other words, our classifier found 32 additional distinct known pulsars that were not included in the benchmark date set. 
	Among these 84 beams, 3 beams contained dim detections of known pulsars that could easily be missed by manual inspection.	
	This would be really problematic if these dim detections were the only observations of unknown pulsars. 
	One such example was shown earlier in Fig~\ref{fig:periocic_pulses};
	\item By comparing our discovery list with the PALFA New Pulsars List, we concluded that our classifier was able to detect all known pulsars (that were found by other single-pulse search approaches) in our data set;
	\item There were a number of dim periodic candidates that could be pulsars. We included the best five of those as PDs for further exploration. One such example is shown in Fig.~\ref{fig:possible_discovery}. It can be difficult to find these dim candidates through manual inspection.
	Furthermore, note that SPEGID identified over 4000 five-SPEG groups from those 1120 beams.
	However, because none of these five-SPEG groups contained any positively classified SPEG (by our trained model), they were classified as UCs and FPs during manual inspection.
	\item In total, we found that our single-pulse search approach was able to find 92 out of the 284 (i.e., 32.4\%) known pulsars (that were discovered in both single-pulse search and periodicity search approaches) in beams that were positioned within $15\arcmin$ of their known locations;
	\item During manual inspection, we also found 361 RRAT-like candidates (SPEG groups with one or two good pulse candidates at the same DM with maximum S/N $>$ 7) from those 1120 beams that were classified as UCs and FPs.
	It is known that RRATs usually have few pulses detected within a short observation period. 
	When the RRAT-like candidates are neither bright enough nor can a period be found among SPEGs, it becomes difficult to confirm their astronomical origin through the two approaches described in Sections \ref{sect:freq-time} and \ref{sect:periodicity}.
	Correspondingly, due to the lack of confirmed examples in the benchmark data set, these RRAT-like candidates were classified as non-pulsars by our classifier. 
\end{enumerate} 

\section{Discussion of the challenges and implications}
\label{sect:discussion}

The automatic approach to radio pulsar detection presented in this paper showed much improvement compared with our previous work \citep{Devine2016} if measured by predictive power. 
More specifically, compared with the best classifier of DPGs, our best classifier of SPEGs exhibited significant increase in F-Measure (from 71.6\% to 96.8\%). (Here F-Measure by SPEG is used because F-Measure by beam was not provided in Devine et al. \citeyear{Devine2016}.)
We believe this is mainly because we correctly identified individual pulses as separate SPEGs in the DM and time space using DBSCAN clustering and merging the clusters.
Therefore, compared with RAPID which identified DPGs in the composite S/N versus DM subplot for an entire observation, we could extract more distinctive features of astrophysical pulses to separate them from RFI and noise. 

Applying DBSCAN clustering to identify related single-pulse event groups in the DM versus time space seems natural and fairly simple at first sight because single-pulse events appear to form denser areas. 
However, in practice, simply applying DBSCAN clustering could end up either grouping unrelated single-pulse events together or splitting related single-pulse events into many clusters because of the following two reasons: 
(i) Pulses have a wide range of time and DM extents, depending on the pulse width, shape, and DM. These different time and DM extents require different cluster radii. 
On the one hand, for dim and/or narrow pulses which usually have a small time and DM extent, including any noise event would have a more significant influence on the overall features of the resultant SPEG, necessitating a smaller radius. 
Furthermore, a small radius is also required to separate pulses that are detected closely in time; 
on the other hand, because bright, wide pulses will be detected over larger time and DM extents than their dim, narrow counterparts, a larger radius is needed to overcome possible large time difference between adjacent single-pulse events;
(ii) Data processing techniques can also change the distance between single-pulse events. For example, clipping can lead to larger distances between remaining single-pulse events; furthermore, the distance likely varies with DM. 
To address these challenges, we used an adequately small radius
and introduced the merging of clusters thereafter.

Note that DBSCAN clustering allowed us to find the single-pulse events that were used to calculate the expected DM and time span in the merging step. 
Moreover, when we used equations~(\ref{eq:SNR_DM}) and~(\ref{eq:SNR_DM_erf}) to calculate the expected DM and time span, we faced the discrepancy between the theoretical S/N decline and the observed S/N decline, and the problem of clipped pulses as well. 
We solved these two challenges by using empirical formulas (equations~(\ref{eq:SNR_thresh}) and~(\ref{eq:SNR_thresh_clipped})) to calculate $\textrm S_\textrm{th}$ (instead of using a fixed value of 5) and obtained satisfactory expected DM and time ranges.  
Thus both DBSCAN clustering and merging the resultant clusters are necessary for the identification of SPEGs.

Furthermore, 
SPEGs may vary significantly in brightness, width and shapes. 
In order to identify the peak-like shape within SPEGs in the S/N versus DM space, we developed a new peak scoring algorithm.
Using a low, DM-spacing-dependent threshold for the Fit-Line-Slope allowed us to detect wide and/or dim pulses, as well as pulses of various shapes. 
However, this also increased the number of non-astrophysical SPEGs.

On the computational side, although DBSCAN clustering, merging the clusters, and peak scoring combined resulted in a significant denoising effect in the DM versus time space, the number of SPEGs that needed to be classified was still high. Therefore we defined the noise level for each beam and only classified relatively bright SPEGs. Furthermore, we excluded the $\chi^{2}$ of the S/N versus DM from our feature list because the fitting takes a long time and does not always converge for every pulse, making it impractical to use on large data sets.

In this paper we selected RandomForest with SMOTE treatment as the best learner because of its very high recall and precision, and very low FPR. It should be noted that all six classifiers had over 93\% recall when SMOTE treatment was applied, but the other five classifiers had significantly lower precision and higher FPR. RandomForest showed the highest precision and lowest FPR most likely because, compared with other learners, it reduced the problem of overfitting. 
The fact that RandomForest with SMOTE was also selected as the best learner for the classification stage in our previous work \citep{Devine2016} (which was based on the GBT Drift-scan survey) provides some generalisability of the finding that RandomForest in combination with SMOTE performs well for automatic pulsar classification. Testing the learners considered in this work and other additional learners on different surveys would help further exploring the generalisability this finding.

A close examination of our results based on the benchmark data set provides some ideas for further improvement of the classification performance. 
We found that three of the four pulsars missed by RandomForest with SMOTE treatment had less than five fairly dim pulses detected, and the fourth pulsar had a very low DM. 
The inductive learning hypothesis states that 
``any hypothesis found to approximate the target function well over a sufficiently large set of training examples will also approximate the target function well over other unobserved examples'' \citep{Mitchell:1997:ML:541177}.
This suggests that our current benchmark data set does not contain enough dim pulsars.
In order to improve the classifier's ability to discover dim pulsars, more dim pulsars (once they are discovered), or marginal detections of bright pulsars (once they are found, as those shown in Figs.~\ref{fig:periocic_pulses} and \ref{fig:possible_discovery}), should be included into the benchmark data set and used for training. 
This agrees with the generally accepted notion that machine learning is not one-shot process of building a data set and applying a learner, but rather an iterative process that includes applying the learner, analysing the results, modifying the data and/or the learner, and repeating \citep{Domingos2012}.
Therefore, we believe that the results can further be improved by creating a more representative benchmark data set, constructing more distinctive features, conducting more hyperparameter optimization, and so on.

The fact that we used periodicity to confirm the astrophysical nature of pulses when three or more pulses were detected at the same DM was helpful for detection of pulsar candidates consisting of a series of fairly dim pulses.
However, it remains challenging to discriminate faint, isolated pulses from noise.

Finally, it is important to emphasize that SPEGID is applicable to pulsar surveys of different observational setups, as long as the survey-specific parameters (i.e., the central observing frequency ($\nu$) and the total bandwidth ($\Delta \nu$)) are modified accordingly.
For example, although in this paper we only presented the results based on data from the PALFA survey, we also tested SPEGID on data from GBT Drift-scan survey (without changing any of the parameters except for $\nu$ and $\Delta \nu$) and received similar performance. 
All one needs is to run the procedures of stages 1 and 2 on the survey in hand, preferably using a benchmark data set with pulsar examples from the same survey to train the learner.
However, as mentioned earlier, the values of parameters to some degree depend on the research goals. For that reason, parameters' values may be slightly modified, whereas changing the forms of equations~(\ref{eq:SNR_DM}), (\ref{eq:SNR_DM_erf}), (\ref{eq:SNR_thresh}) and (\ref{eq:SNR_thresh_clipped}) is not necessary.

\section{Conclusions}
\label{sect:conclusion}
We developed a novel, automatic two stage single-pulse search approach in order to improve the speed and accuracy of single-pulse search analysis.
In the first stage, SPEGID identified astrophysical pulse candidates as SPEGs through clustering the single-pulse events and merging the clusters. In addition, a new peak scoring algorithm was applied to SPEGs to identify their associated peaks in the S/N and DM space. 
SPEGs that showed up at a consistent DM were then grouped together.
In the second stage, we used supervised machine learning approach to classify SPEGs to pulsar and non-pulsar classes. For that purpose, 
a benchmark data set containing 60 representative pulsars 
(detected in 90 beams)
was created.
Our machine learning experiments on the benchmark data set showed that the classifier based on RandomForest with SMOTE imbalance treatment had the best performance with respect to F-Measure. 

The unlabelled full data set consisting of 47,042 independent beams that are observations of 33,536 unique positions from the PALFA survey was processed using SPEGID to identify SPEGs. Their features were then extracted, so these SPEGs could be automatically classified using the best learner trained on the whole benchmark data set. 24,426 positively classified SPEGs that belonged to 1,299 beams were examined manually. 
The results showed that our model detected all 60 known pulsars (in 90 beams) included in the benchmark data set, as well as 32 additional known pulsars (in 84 beams) not included in the benchmark data set, among which several were fairly dim.  
Additionally, it also found five dim pulsar candidates with multiple periodic signals that are worth further exploration.
Our future work will be focused on building a larger benchmark data set that includes more dim pulsars and applying multi-class classification on it.

Developing methods for automatic classification of single-pulse search output is necessary if astronomers are to cope with the demands of sensitive surveys with CHIME, MeerKAT, and the FAST telescope. CHIME is expected to result in the discovery of dozens or more bright FRBs each day. Developing methods for automatic classification will be necessary to probe the low-intensity end of this population to truly understand the FRB luminosity distribution. The FAST telescope is expected to operate in multi-beam drift-scan mode for the first several years of operations, with extremely short ($\sim$seconds) observation lengths. This means that a large fraction of pulsars will only be detectable through their single pulse emission, necessitating intelligent algorithms for single-pulse searches to analyse the large amount of survey data produced. In addition to discovering sporadic pulsars and FRBs, and placing limits on the low-intensity members of those populations, intelligent algorithms for single-pulse searches have the potential to recognize completely new source classes. Future work will explore different regimes of search space (i.e., broader pulses, very low DMs) in order to perhaps discover other populations of radio-emitting objects.
\section*{Acknowledgements}
\label{sect:ack}
This work is partially supported by the National Science Foundation under Award No. OIA-1458952. We thank the PALFA collaboration for making these data available and for discussions that improved this paper.
We also thank the editor and the reviewer for their valuable comments.




\bibliographystyle{mnras}
\bibliography{single_pulse_search} 

\begin{thebibliography}{}
\makeatletter
\relax
\def\mn@urlcharsother{\let\do\@makeother \do\$\do\&\do\#\do\^\do\_\do\%\do\~}
\def\mn@doi{\begingroup\mn@urlcharsother \@ifnextchar [ {\mn@doi@}
  {\mn@doi@[]}}
\def\mn@doi@[#1]#2{\def\@tempa{#1}\ifx\@tempa\@empty \href
  {http://dx.doi.org/#2} {doi:#2}\else \href {http://dx.doi.org/#2} {#1}\fi
  \endgroup}
\def\mn@eprint#1#2{\mn@eprint@#1:#2::\@nil}
\def\mn@eprint@arXiv#1{\href {http://arxiv.org/abs/#1} {{\tt arXiv:#1}}}
\def\mn@eprint@dblp#1{\href {http://dblp.uni-trier.de/rec/bibtex/#1.xml}
  {dblp:#1}}
\def\mn@eprint@#1:#2:#3:#4\@nil{\def\@tempa {#1}\def\@tempb {#2}\def\@tempc
  {#3}\ifx \@tempc \@empty \let \@tempc \@tempb \let \@tempb \@tempa \fi \ifx
  \@tempb \@empty \def\@tempb {arXiv}\fi \@ifundefined
  {mn@eprint@\@tempb}{\@tempb:\@tempc}{\expandafter \expandafter \csname
  mn@eprint@\@tempb\endcsname \expandafter{\@tempc}}}

\bibitem[\protect\citeauthoryear{Bethapudi \& Desai}{Bethapudi \&
  Desai}{2018}]{Bethapudi2018}
Bethapudi S.,  Desai S.,  2018, \mn@doi [Astronomy and Computing]
  {10.1016/j.ascom.2018.02.002}, 23, 15

\bibitem[\protect\citeauthoryear{Breiman}{Breiman}{2001}]{Breiman2001}
Breiman L.,  2001, \mn@doi [Machine Learning] {10.1023/A:1010933404324}, 45, 5

\bibitem[\protect\citeauthoryear{Burke-Spolaor \& Bailes}{Burke-Spolaor \&
  Bailes}{2010}]{Burke-Spolaor2010}
Burke-Spolaor S.,  Bailes M.,  2010, \mn@doi [\mnras]
  {10.1111/j.1365-2966.2009.15965.x}, 402, 855

\bibitem[\protect\citeauthoryear{Chawla, Bowyer, Hall  \& Kegelmeyer}{Chawla
  et~al.}{2002}]{chawla2002smote}
Chawla N.~V.,  Bowyer K.~W.,  Hall L.~O.,   Kegelmeyer W.~P.,  2002, Journal of
  artificial intelligence research, 16, 321

\bibitem[\protect\citeauthoryear{Cohen}{Cohen}{1995}]{Ripper}
Cohen W.~W.,  1995, in Twelfth International Conference on Machine Learning.
  Morgan Kaufmann, pp 115--123, \mn@doi{10.1016/B978-1-55860-377-6.50023-2}

\bibitem[\protect\citeauthoryear{Cordes \& McLaughlin}{Cordes \&
  McLaughlin}{2003}]{Cordes2003}
Cordes J.~M.,  McLaughlin M.~A.,  2003, \mn@doi [\apj] {10.1086/378231}, 596,
  1142

\bibitem[\protect\citeauthoryear{Cordes et~al.,}{Cordes
  et~al.}{2006}]{Cordes2006}
Cordes J.~M.,  et~al., 2006, \mn@doi [\apj] {10.1086/498335}, 637, 446

\bibitem[\protect\citeauthoryear{Deneva et~al.,}{Deneva
  et~al.}{2009}]{Deneva2009}
Deneva J.~S.,  et~al., 2009, \mn@doi [\apj] {10.1088/0004-637X/703/2/2259},
  703, 2259

\bibitem[\protect\citeauthoryear{{Devine}, {Goseva-Popstojanova}  \&
  {McLaughlin}}{{Devine} et~al.}{2016}]{Devine2016}
{Devine} T.~R.,  {Goseva-Popstojanova} K.,   {McLaughlin} M.,  2016, \mn@doi
  [\mnras] {10.1093/mnras/stw655}, \href
  {http://adsabs.harvard.edu/abs/2016MNRAS.459.1519D} {459, 1519}

\bibitem[\protect\citeauthoryear{Domingos}{Domingos}{2012}]{Domingos2012}
Domingos P.,  2012, \mn@doi [Communications of the ACM]
  {10.1145/2347736.2347755}, 55, 78

\bibitem[\protect\citeauthoryear{Ester, Kriegel, Sander  \& Xu}{Ester
  et~al.}{1996}]{Ester1996}
Ester M.,  Kriegel H.~P.,  Sander J.,   Xu X.,  1996, \mn@doi [Second
  International Conference on Knowledge Discovery and Data Mining]
  {10.1.1.71.1980}, pp 226--231

\bibitem[\protect\citeauthoryear{Frank \& Witten}{Frank \& Witten}{1998}]{PART}
Frank E.,  Witten I.~H.,  1998, in Shavlik J.,  ed., Fifteenth International
  Conference on Machine Learning. Morgan Kaufmann, pp 144--151

\bibitem[\protect\citeauthoryear{Frank, Hall  \& Witten}{Frank
  et~al.}{2016}]{Weka}
Frank E.,  Hall M.~A.,   Witten I.~H.,  2016, The WEKA Workbench. Online
  Appendix for "Data Mining: Practical Machine Learning Tools and Techniques".
Morgan Kaufmann, Fourth Edition

\bibitem[\protect\citeauthoryear{G{\'o}mez \& Rojas}{G{\'o}mez \&
  Rojas}{2016}]{Gomez2016}
G{\'o}mez D.,  Rojas A.,  2016, \mn@doi [Neural Computation] {10.1162/NECO},
  28, 216

\bibitem[\protect\citeauthoryear{Han, Kamber  \& Pei}{Han
  et~al.}{2011}]{Han:2011:DMC:1972541}
Han J.,  Kamber M.,   Pei J.,  2011, Data Mining: Concepts and Techniques, 3rd
  edn.
Morgan Kaufmann Publishers Inc., San Francisco, CA, USA

\bibitem[\protect\citeauthoryear{Karako-Argaman et~al.,}{Karako-Argaman
  et~al.}{2015}]{Karako-Argaman2015}
Karako-Argaman C.,  et~al., 2015, \mn@doi [\apj] {10.1088/0004-637X/809/1/67},
  809, 67

\bibitem[\protect\citeauthoryear{Keith, Eatough, Lyne, Kramer, Possenti, Camilo
   \& Manchester}{Keith et~al.}{2009}]{Keith2009}
Keith M.~J.,  Eatough R.~P.,  Lyne A.~G.,  Kramer M.,  Possenti A.,  Camilo F.,
    Manchester R.~N.,  2009, \mn@doi [\mnras]
  {10.1111/j.1365-2966.2009.14543.x}, 395, 837

\bibitem[\protect\citeauthoryear{Lazarus et~al.,}{Lazarus
  et~al.}{2015}]{Lazarus2015}
Lazarus P.,  et~al., 2015, \mn@doi [\apj] {10.1088/0004-637X/812/1/81}, 812, 81

\bibitem[\protect\citeauthoryear{{Lorimer} \& {Kramer}}{{Lorimer} \&
  {Kramer}}{2004}]{2004hpa..book.....L}
{Lorimer} D.~R.,  {Kramer} M.,  2004, {Handbook of Pulsar Astronomy}.
Cambridge Univ. Press, Cambridge

\bibitem[\protect\citeauthoryear{Lorimer, Bailes, McLaughlin, Narkevic  \&
  Crawford}{Lorimer et~al.}{2007}]{Lorimer2007}
Lorimer D.~R.,  Bailes M.,  McLaughlin M.~A.,  Narkevic D.~J.,   Crawford F.,
  2007, \mn@doi [Science] {10.1126/science.1147532}, 318, 777

\bibitem[\protect\citeauthoryear{{Manchester}, {Hobbs}, {Teoh}  \&
  {Hobbs}}{{Manchester} et~al.}{2005}]{ATNF_catalog}
{Manchester} R.~N.,  {Hobbs} G.~B.,  {Teoh} A.,   {Hobbs} M.,  2005, \mn@doi
  [\aj] {10.1086/428488}, \href
  {http://adsabs.harvard.edu/abs/2005AJ....129.1993M} {129, 1993}

\bibitem[\protect\citeauthoryear{McLaughlin et~al.,}{McLaughlin
  et~al.}{2006}]{McLaughlin2006}
McLaughlin M.~A.,  et~al., 2006, \mn@doi [Nature] {10.1038/nature04440}, 439,
  817

\bibitem[\protect\citeauthoryear{Mitchell}{Mitchell}{1997}]{Mitchell:1997:ML:541177}
Mitchell T.~M.,  1997, Machine Learning, 1 edn.
McGraw-Hill, Inc., New York, NY, USA

\bibitem[\protect\citeauthoryear{Murphy}{Murphy}{2012}]{Murphy:2012:MLP:2380985}
Murphy K.~P.,  2012, Machine Learning: A Probabilistic Perspective.
The MIT Press

\bibitem[\protect\citeauthoryear{Pedregosa et~al.,}{Pedregosa
  et~al.}{2011}]{Scikit-learn}
Pedregosa F.,  et~al., 2011, \mn@doi [J. Mach. Learn. Res.]
  {10.1007/s13398-014-0173-7.2}, 12, 2825

\bibitem[\protect\citeauthoryear{Pham \& Triantaphyllou}{Pham \&
  Triantaphyllou}{2008}]{Pham2008}
Pham H. N.~A.,  Triantaphyllou E.,  2008, The Impact of Overfitting and
  Overgeneralization on the Classification Accuracy in Data Mining.
Springer US, Boston, MA, pp 391--431, \mn@doi{10.1007/978-0-387-69935-6_16},
  \url {https://doi.org/10.1007/978-0-387-69935-6_16}

\bibitem[\protect\citeauthoryear{Platt}{Platt}{1999}]{Platt:1999}
Platt J.~C.,  1999, MIT Press, Cambridge, MA, USA, Chapt. Fast Training of
  Support Vector Machines Using Sequential Minimal Optimization, pp 185--208,
  \url {http://dl.acm.org/citation.cfm?id=299094.299105}

\bibitem[\protect\citeauthoryear{Quinlan}{Quinlan}{1993}]{Quinlan:1993:CPM:583200}
Quinlan J.~R.,  1993, C4.5: Programs for Machine Learning.
Morgan Kaufmann Publishers Inc., San Francisco, CA, USA

\bibitem[\protect\citeauthoryear{{Ransom}}{{Ransom}}{2001}]{RansomPhDT}
{Ransom} S.~M.,  2001, PhD thesis, Harvard Univ.

\bibitem[\protect\citeauthoryear{Rasmussen}{Rasmussen}{1992}]{Rasmussen1992}
Rasmussen E.,  1992, Clustering Algorithms. In: Frakes WB, Baeza-Yates R,
  editors. Information retrieval data structures and algorithms.
Prentice Hall, Englewood Cliffs, NJ, USA

\bibitem[\protect\citeauthoryear{Spitler et~al.,}{Spitler
  et~al.}{2014}]{Spitler2014}
Spitler L.~G.,  et~al., 2014, \mn@doi [\apj] {10.1088/0004-637X/790/2/101},
  790, 101

\bibitem[\protect\citeauthoryear{Swiggum et~al.,}{Swiggum
  et~al.}{2014}]{Swiggum2014}
Swiggum J.~K.,  et~al., 2014, \mn@doi [\apj] {10.1088/0004-637X/787/2/137},
  787, 137

\bibitem[\protect\citeauthoryear{Witten, Frank  \& Hall}{Witten
  et~al.}{2011}]{Witten2011}
Witten I.~H.,  Frank E.,   Hall M.~A.,  2011, Data Mining: Practical Machine
  Learning Tools and Techniques, 3rd edn.
Morgan Kaufmann Publishers Inc., San Francisco, CA, USA

\bibitem[\protect\citeauthoryear{Wolpert}{Wolpert}{1996}]{Wolpert1996}
Wolpert D.~H.,  1996, \mn@doi [Neural Computation]
  {10.1162/neco.1996.8.7.1341}, 8, 1341

\makeatother
\end{thebibliography}




\appendix

\section{Approach to searching for an underlying periodicity among single-pulse events}
\label{sect:appendix}
In Section~\ref{sect:periodicity} we mentioned that our approach to finding an underlying periodicity among a group of SPEGs was based on looking for a common divisor among the gaps of their arrival times. Here we describe this approach in detail.  
Given a group of SPEGs, we search for an underlying periodicity among them in the following way:

(i) The underlying periodicity is searched among single-pulse events that belong to any of the SPEGs;

(ii) The underlying periodicity is searched in the DM channel in which the S/N of the brightest SPEG (within the group) peaks;

(iii) The underlying periodicity is searched only when there are $n \ge 3$ single-pulse events that satisfy the previous two constraints;

(iv) Suppose these $n$ single-pulse events are detected at $t_\textrm{1}$, $t_\textrm{2}$, ..., $t_\textrm{n}$ respectively, first, we find the time difference  $\Delta t_\textrm{i, i+1}$ between two neighboring single-pulse events as follows:
\begin{equation}
\Delta t_\textrm{i, i+1} = t_\textrm{i+1} - t_\textrm{i}\quad (1 \le \textrm{i} < \textrm{n}),
\label{eq:t_ij}
\end{equation}
then we find the minimum time difference:
\begin{equation}
\Delta t_\textrm{min} = \textrm{min} (\Delta t_\textrm{i, i+1});
\label{eq:t_min}
\end{equation}

(v) Once $\Delta t_\textrm{min}$ is determined, a series of period candidates ($T_\textrm{k}$) can be obtained by equation~(\ref{eq:T_search}):
\begin{equation}
T_\textrm{k} = \frac{\Delta t_\textrm{min}}{\textrm{k}} \quad (k = 1, 2, 3, ...),
\label{eq:T_search}
\end{equation}
and all period candidates no less than 0.05 s are searched;

(vi) To determine whether a period candidate $T_\textrm{k}$ is valid for two neighboring single-pulse events, we introduce the concept of the tolerance level ($\alpha$) and compare it with $\beta$, the remainder of $\Delta t_\textrm{i, i+1}$ divided by $T_k$:
\begin{equation}
\beta = \big(\big(\frac{\Delta t_\textrm{i, i+1}}{T_k}\big)\bmod 1 \big).
\label{eq:tolerance}
\end{equation}
If $\beta$ < $\alpha$, or $\beta$ > 1 - $\alpha$, then this period candidate $T_\textrm{k}$ between these two events is considered valid. 
$\alpha$ is usually set between 0.01  and 0.05, and lower $\alpha$ value indicates more accurate periodicity;

(vii) If the period candidate $T_\textrm{k}$ is valid for all two neighboring event pairs, then an underlying periodicity (with a period of $T_\textrm{k}$) is found among these $\textrm{n}$ single-pulse events. The searching is continued until $T_\textrm{k}$ is less than 0.05 s;

(viii) If none of the period candidates is valid for these $\textrm{n}$ events, then we search for an underlying periodicity in all subsets with $\textrm{n} - 1 $ events using the same procedure described from (iii) to (vii);

(ix) The searching stops either when there are less than three events within the group, or when a valid period is found among all single-pulse events within the group.


\bsp	
\label{lastpage}
\end{document}